\documentclass[%
 reprint,
 amsmath,amssymb,
 aps,
]{revtex4-2}

\usepackage{graphicx}
\usepackage{dcolumn}
\usepackage{bm}
\usepackage{xcolor}

\begin{document}

\preprint{APS/DiracTunneling}

\title[Superluminal tunneling times without superluminal signaling]{Superluminal tunneling times without superluminal signaling: Fading of the MacColl-Hartman effect at early times}

\author{Randall S. Dumont}
\affiliation{Department of Chemistry and Chemical Biology,\\McMaster University, Hamilton, Ontario, Canada L8S 4M1}
\email{dumontr@mcmaster.ca}

\author{Tom Rivlin}
\affiliation{Chemical and Biological Physics Department,\\Weizmann Institute of Science, 
 76100 Rehovot, Israel}
 \email{Present address: Atominstitut, TU Wien, 1020 Vienna, Austria}
\email{tom.rivlin@tuwien.ac.at}
\email{tom.rivlin@weizman.ac.il}
\date{\today}

\begin{abstract}
A curious feature of quantum tunneling known as the MacColl-Hartman effect results in the numerical observation that particles can traverse a barrier with effective superluminal speed. However, because tunneling is never certain, any attempt to use this effect to send a signal faster than light would require sending many particles.  In this work, we consider sending---in parallel, without interactions between particles---sufficiently many particles to ensure at the least one of them tunnels.  In this case, in spite of the time advance of the mean time for a single tunneling particle, the mean time to send one bit of information is larger for tunneling particles than for the same number of free photons. This removes any possibility of superluminal signaling.  We show that the mean time to send one bit using $N$ particles is determined by the early-time tail of the distribution of tunneling times for one particle and that, when this early-time tail is highly accurately modeled using steepest descent, the MacColl-Hartman effect is seen to fade away.
\end{abstract}

%


\maketitle

\section{Introduction}

Means of measuring quantum tunneling time have been debated for decades, with renewed interest in recent years \cite{84PollakMiller, 89HaugeStovneng, 91LowMende, 92MugaBrouard, 93Leavens, 94BrouardSala, 95Steinberga, 95Steinbergb, 97Hauge, 00MugaLeavens, 01Ruseckas, 01RuseckasKaulakys, 03Winful, 08Muga, 08EgusquizaMuga, 08MayatoAlonso, 08Schulman, 08SokolovskiMuga, 12Galapon, 15McDonaldOrlando, 20PablicoGalapon, 20DumontRivlinPollak, 21RivlinPollakDumont, 21bRivlinPollakDumont, 21IanconescuPollak, 21Sokolovski}. In some interpretations, the time for a wavepacket to traverse a barrier can be faster than the time it would take an equivalent photon wavepacket traveling through free space. This results from the MacColl-Hartman effect \cite{32Maccoll, 62Hartman}, which says the \textit{phase time} of tunneled particles is independent of barrier width for sufficiently wide barriers---theoretically allowing for arbitrarily fast speeds. 

In previous works \cite{20DumontRivlinPollak, 21RivlinPollakDumont, 21bRivlinPollakDumont}, we assessed the problem relativistically with the Dirac equation and non-relativistically with the Schr\"{o}dinger equation and verified the effect. We also showed the relationship between the phase time, the flight time, and other tunneling times. In contrast to attosecond experiments that purport to find instantaneous, zero-time tunneling \cite{19SainadhXu}, our previous works demonstrated that tunneling times are short but finite (similar to the experimental findings of, e.g., Ref. \cite{20SpieringsSteinberg}). We further showed that the phase time is the appropriate way of measuring the barrier interaction time and it can lead to faster-than-light times. Additionally, we demonstrated that ``momentum filtering''---the disproportionate selection of higher-momentum components of the wave packet due to tunneling---cannot be used to explain away this tunneling-related speedup, as the speedup occurs even when controlling for momentum filtering. Lastly, we dismissed arguments relating to a lack of ``causal'' connection between the pre- and post-tunnel wavepacket by showing that the tunneled wave packet's features depended heavily on the initial wave packet.

The controversial result \cite{94SpielmannSzipocs, 96Diener, 97ChiaoSteinberg, 03Nimtz, 06Winful, 06Winfulb, 17NimtzAichmann} we supplied was supported by an argument for why the faster-than-light times do not imply superluminal signaling, let alone violations of causality: the drastic loss of amplitude of the tunneled wave packet more than compensates for the small ``speedup,'' meaning nontunneled photons are always preferable for sending signals as quickly as possible. 

That argument relied on steepest-descent approximations (SDAs) \cite{93DumontMarchioro, 07DeLeoRotelli} of the evolving wave function at the most probable transmission time. Other times were treated with Taylor expansions about that time. This work seeks to make our previous conclusions more rigorous by clarifying what constitutes superluminal signaling and drawing attention to early-time portions of the transmission time probability distribution.  Modeling the distribution at early times leads to a time-dependent SDA which reveals a previously unknown phenomenon: the MacColl-Hartman effect effectively fades away at early times far away from the most probable time---the very times that are key to the question of superluminal signaling. At short times---the fastest arrival times for the transmitted wavepacket---probability distributions nearly match the leading portions of free particle wavepackets traveling at the initial subluminal velocity. While this result is based on SDAs, it is supported by numerical calculations which are exquisitely accurate at these times.  While the time scales required to test these results experimentally are still out of reach, the underlying result that the MacColl-Hartman effect fades away at early times---relevant in the many-particle limit---is in principle testable with simple experimental designs, similar to time-of-flight experiments such as  \cite{10FuhrmanekLanceTuchendler, 15DuLiWen} and ones presented recently in \cite{20SpieringsSteinberg, 20RamosSpierings, 21SpieringsSteinberg}, which follow a long line of inconclusive tunneling time experiments \cite{93SteinbergKwiatChiao, 01LonghiMarano, 12ShafirSoiferBruner, 14LandsmanWeger, 15LandsmanKeller, 15TorlinaMorales, 15PedatzurOrensteinSerbinenko, 16Texier, 19SainadhXu, 20SainadhSangLitvinyuk, 20Kheifets}.

We compute the probability distribution for the tunneled flux as a function of time \cite{17PetersenPollak, 17Pollak, 18PollakMiret} for the understudied subject \cite{02LiChen, 07DeLeoRotelli, 07LunardiManzoni, 13DeLeo, 15IvanovKim} of a massive particle such as an electron traveling at near light speed. In order to compare fast signal transmission when one probability distribution is normalized (the distribution of photon arrival times) while the other is not (the distribution of tunneled electron times) we must consider many parallel processes.  Optimal signal transmission with $N$ particles is characterized by the ``\textit{first-click distribution}," the probability distribution for the earliest signal transmission time among $N$ particles.  Analysis of the first-click distribution leads to the vanishing MacColl-Hartman effect and the impossibility of superluminal signaling using the effect.

\section{Dirac Tunneling}

Tunneling time has been studied extensively by ourselves and others in the nonrelativistic regime using the Schr\"{o}dinger equation.  The relativistic regime we consider here has very similar physics and much of the intuition from one regime carries over into the other (the observables we use are defined in the same way in both cases, for instance).  The principal difference is the dispersion relationship which morphs from quadratic to linear as velocity approaches $c$.  This manifests as slower wave packet broadening.  The time for broadening scales as $\gamma^2$, where $\gamma = \frac{1}{\sqrt{1 - v^{2}/c^{2}}}$ is the ubiquitous Lorentz factor.  Also, the tunneling momentum filtering effect---see below---is smaller than the equivalent for nonrelativistic tunneling.

The time evolution of relativistic electrons is governed by the Dirac equation.  The (time-dependent) Dirac equation has the following form in 1+1 dimensions:
\begin{equation}
i\hbar \frac{\partial }{\partial t}\left(
\begin{array}{c}
\psi _{0} \\
\psi _{1}%
\end{array}%
\right) =\mathsf{H}\left(
\begin{array}{c}
\psi _{0} \\
\psi _{1}%
\end{array}%
\right) ,
\end{equation}
where symbols have their usual meanings in relativistic quantum mechanics, and %
\begin{equation}
\mathsf{H=}mc^{2}\left(
\begin{array}{cc}
1 & 0 \\
0 & -1%
\end{array}%
\right) +ic\hbar \left(
\begin{array}{cc}
0 & 1 \\
1 & 0%
\end{array}%
\right) \frac{\partial }{\partial z}+V\left( z\right) .
\end{equation}
This follows from the $\left(3+1\right)$-dimensional Dirac equation when the potential depends only on $z$; the $x$ and $y$ degrees of freedom separate and the two components here correspond to the conserved spin state of spin up. Flux in the $z$ direction at $z_2$, the observation point to the right of the barrier, is%
\begin{equation}
\mathsf{J} =-\frac{i}{\hbar }\left[ \mathsf{H},\Theta _{\left( z_{2},\infty \right) }\right]
= c \left(
\begin{array}{cc}
0 & 1 \\
1 & 0%
\end{array}%
\right) \delta \left( z-z_{2}\right) ,
\end{equation}
where $\Theta _{\left( z_{2},\infty \right)}$ is the step function which equals 1 inside the interval and zero otherwise.  The time-dependent flux, which is specific to the initial wave packet, is the expectation value of $\mathsf{J}$:%
\begin{equation}
P\left(t\right)= 2c\, {\rm{Re}}\left( \psi^{\ast}_{0}\left(z_{2},t\right) \psi_{1}\left(z_{2},t\right)\right).
\end{equation}

We consider a single rectangular barrier: $V(z)=V_{\rm{top}}$ for $z_{1}<z<z_{2}$ and  $V(z)=0$ otherwise. The rectangular barrier provides the best-case scenario to see possible superluminal signaling, as all of the wavepacket components with energy less than the barrier top (plus the particle rest energy) must tunnel across the entire width of the barrier.  For a smooth potential such as a Gaussian barrier, the tunneling gap gets smaller for wavepacket components with energy approaching the above-barrier threshold \cite{21bRivlinPollakDumont}.  These components consequently have a diminishing MacColl-Hartman effect.

Here we expand on and refine a similar method introduced in \cite{20DumontRivlinPollak}. To compute the time-evolving wavepacket, we transform from the energy representation,
\begin{equation}
\mathbf{\psi }\left( z,t\right) =\int_{-\infty }^{\infty }dE\,b_{E}\exp
\left( \frac{-iEt}{\hbar }\right) \mathbf{\psi }_{E}\left( z\right),
\end{equation}
where the expansion coefficient, $b_{E}$, is determined from the initial wavepacket.
In terms of momentum,
\begin{equation}
p\left( E\right) =\sqrt{E^{2}/c^{2}-m^{2}c^{2}},
\end{equation}
the integral representation of the evolving wavepacket takes on a simple form if the initial wavepacket is a Gaussian (with a spinor factor) far to the left of the barrier.  Specifically,
\begin{multline}
\mathbf{\psi}(z,t) = K\int_{-\infty }^{\infty } dp \exp \left( -
\frac{\left( p-p_{0}\right) ^{2}}{2\Gamma }\right) T(p)
\mathbf{u}(p)\times \\ \exp \left( i\left(\frac{p\left(
z-z_{0}\right) -E\left( p\right) t}{\hbar}\right)\right),
\label{psi_integral_over_p}
\end{multline}
where
$K$ is a normalization factor and
\begin{equation}
\mathbf{u}( p) =\left(
\begin{array}{c}
1 \\
\frac{cp}{E\left( p\right) +mc^{2}}%
\end{array}%
\right)
\end{equation}
is a two-component spinor.  The transmission amplitude for a single
barrier has the form \cite{07DeLeoRotelli}%
\begin{equation}
T(p) =\frac{\exp \left( -ipl/\hbar \right) }{\cosh \left(
ql/\hbar \right) +\frac{1+\alpha ^{2}}{2\alpha }\sinh \left( ql/\hbar
\right) },
\end{equation}
where%
\begin{equation}
\alpha =i\frac{q}{p}\left(\frac{E+m}{E-V_{\rm{top}}+m}\right),
\end{equation}
\begin{equation}
q = \sqrt{m^{2}c^{2}-\left(E-V_{\rm{top}}\right)^{2}/c^{2}}
\end{equation}
and $l=z_{2}-z_{1}$ is the width of the barrier.

The evolving wave packet is computed very accurately below using Eq. (\ref{psi_integral_over_p}) with a nonuniform momentum grid with grid point density chosen to remove cusps in the integrand associated with critical points (where $q=0$). Note that we carefully choose the initial wave packet position such that there is negligible ($\sim 10^{-30}$) initial amplitude inside the barrier and we choose its width such that there is negligible above-barrier transmission and such that momentum filtering is negligible.  We also compute the evolving wave packet using the SDA.  Specifically, $\mathbf{\psi }\left( z,t\right) \sim \mathbf{\psi_{\rm{sd}} }\left( z,t\right)$,
where
\begin{equation}
\mathbf{\psi_{\rm{sd}} }\left( z,t\right) = K\frac{\left( 2\pi \right) ^{1/2}}{\left\vert \frac{\partial ^{2}F}{%
\partial p^{2}}\left( p^{\sharp },t\right) \right\vert ^{1/2}}\mathbf{u}%
\left( p^{\sharp }\right) \exp \left( \frac{i}{\hbar }%
F(p^{\sharp },t) \right),
\end{equation}
\begin{equation}
F(p,t) =-i\hbar \left(-\frac{\left( p-p_{0}\right) ^{2}}{2\Gamma }+\ln\left(T(p)\right)\right) -E(p) t,
\end{equation}
and $p^{\sharp }$ solves%
\begin{equation}
\frac{p-p_{0}}{\Gamma }-\frac{d\ln\left( T(p)\right)}{dp}+\frac{i}{\hbar }\frac{dE(p)}{dp}t=0,
\end{equation}
or%
\begin{equation}
\frac{p-p_{0}}{\Gamma }=\frac{i}{\hbar }\left(\tau(p)-t\right) v(p),
\label{psharp}
\end{equation}
in terms of velocity, $v\left(p\right)$, and the complex phase time,
\begin{equation}
\label{eq:phasetimedef}
\tau(p) = -i\hbar \frac{d\ln\left(T(p\left(E\right))\right)}{dE} .
\end{equation}
Henceforth $\sharp$-superscripted variables are evaluated at $p^\sharp$. 

Here, the SDA is an asymptotic approximation corresponding to the $\hbar \rightarrow 0$ limit. This means the accuracy of the approximation depends on action along the integration path being large compared with $\hbar$.  This accuracy is verified below by comparison with very accurate numerical calculations.

In our previous work, Eq. (\ref{psharp}) was solved for $t=t_{\rm{mp}}$, the time for which the tunneling time distribution in the SDA, $P_{\rm{sd}}(t)$, is a maximum.  If $P_{\rm{sd}}(t)$ is normalized to integrate to 1, it becomes the postselected transmission time distribution.  Since $P_{\rm{sd}}(t)$ is approximately symmetrical about its maximum, $t_{\rm{mp}}$ is approximately the mean postselected transmission time in the SDA.  $F( p^{\sharp },t)$ was evaluated for other times by expanding to second order in $t-t_{\rm{mp}}$.  The resulting ``frozen'' SDA to the tunneling time distribution has the form
\begin{multline}
P_{\rm{sd0}}\left( t\right) = C_{\rm{trans,sd0}}\frac{v^{\sharp }_{\rm{mp}}}{\left(\pi\Gamma\right)^{1/2}\left|\Delta^{\sharp }_{\rm{mp}}\right|}\times \\
\exp\left(-\frac{v^{\sharp }_{\rm{mp}} \,^{2}}{\Gamma\left|\Delta^{\sharp }_{\rm{mp}}\right|^{2}}\left(t-t_{\rm{mp}}\right)^{2}\right),%
\label{Psd0}
\end{multline}
where $\Delta^{\sharp }_{\rm{mp}}=\Delta\left(p^{\sharp }_{\rm{mp}}\right)$,
\begin{equation}
\Delta\left(p\right)=i\frac{\hbar}{\Gamma}+\frac{d}{dp}\left(v\left(p\right)\tau\left(p\right)\right)-\frac{t_{\rm{mp}}}{m\gamma^{2}},
\end{equation}
and
\begin{equation}
C_{\rm{trans,sd0}} = \frac{\left(E_{0}+mc^{2}\right)\gamma^{\sharp }_{\rm{mp}}}{\left(E^{\sharp }_{\rm{mp}}+mc^{2}\right)\gamma_{0}}\exp\left(-\frac{2}{\hbar} {\rm{Im}}
\left( F(p^{\sharp}_{\rm{mp}},t_{\rm{mp}})\right)\right)
\label{Csd0}
\end{equation}
is the associated transmission probability approximation.  Here, variables subscripted by ``mp'' are evaluated at the most probable time.  Equations (\ref{Psd0}) and (\ref{Csd0}) were found to closely match accurate numerical computations, as long as the energy of the incoming wave packet was well below $V_{\rm{top}}+mc^{2}$.  Otherwise, there is an additional above-barrier component. Note that we also choose wave packet initial energies such that ``Klein zone'' \cite{99NittaKudoMinowa} contributions and pair-production effects are negligible.

Note that, when evaluated at $t_{\rm{mp}}$, the real component of $\tau^{\sharp}$ corresponds to the Wigner phase time \cite{89HaugeStovneng, 16Texier} and the imaginary component corresponds to the Pollak-Miller imaginary time \cite{84PollakMiller}. In \cite{21RivlinPollakDumont}, the Wigner phase time was shown to be the tunneling time, but here we generalize these concepts to functions that vary with physical time $t$. At times other than the most probable, the real and imaginary parts of $\tau^{\sharp}$ correspond to ``effective'' phase times and Pollak-Miller times, and hence the properties of these effective times may differ when evaluated at earlier or later physical times.

The Gaussian approximation to the tunneling time distribution provided by Eq. (\ref{Psd0}) manifests the MacColl-Hartman effect via the expression for the most probable time.  The condition for the most probable time is the condition for ${\rm{Im}} F\left(p^{\sharp},t\right)$ to be a minimum.  This occurs when $p^{\sharp}$ is real.  In this case, $t={\rm{Re}} \left(\tau\right)$.  Specifically,
\begin{equation}
t_{\rm{mp}} = t^{\sharp }_{\rm{mp}}\left(z_{0},z_{1}\right)+\sigma^{\sharp }_{\rm{mp}},
\label{tmp}
\end{equation}
where $t^{\sharp }_{\rm{mp}}\left(z_{0},z_{1}\right)=\left(z_{1}-z_{0}\right)/v^{\sharp }_{\rm{mp}}$ is the time required to travel from $z_{0}$ to the left edge of the barrier, $z_{1}$, at speed, $v^{\sharp }_{\rm{mp}}$. Because velocity is highly insensitive to shifts in momentum in the relativistic regime, this speed is almost the same as the initial speed of the wave packet.  The extra term, $\sigma^{\sharp }_{\rm{mp}}$, can be viewed as an underbarrier contribution to the mean time.  Most notable is that $\sigma^{\sharp }_{\rm{mp}}$ quickly asymptotes to a constant with increasing barrier width.  This is the MacColl-Hartman effect.  Some attosecond experiments have suggested that a tunneling electron spends no time under the barrier, which corresponds to leaving out the $\sigma^{\sharp }_{\rm{mp}}$ term above.  While $\sigma^{\sharp }_{\rm{mp}}$ is small, it is not zero and is evident in numerical results.


\section{The First-Click Distribution} 

Previously we showed that the tunneling time probability distribution, $P(t)$, peaks earlier than the corresponding free photon arrival time distribution, $P_{\gamma}(t)$.  We claimed this does not imply superluminal signaling because $P(t) < P_{\gamma}(t)$ for all $t$, for all cases computed.  Here, we make this connection more explicit.

Suppose we wish to send an alert as quickly as possible, and are successful if at least a single particle reaches our destination---the message is received and interpreted as a binary 1 if the particle is detected at $z_{2}$.  To compensate for the low transmission probability, $C_{\rm{trans}}$, we send a large number $N$ of particles (independently and in parallel). Receipt of the signal is ensured only if $N$ is much larger than $1/C_{\rm{trans}}$.  The time scale for receipt of the signal is characterized by what we call the ``first-click'' distribution, the probability distribution for the first time at which one of the $N$ particles is detected at $z_{2}$ (the first ``click'' the detector makes), noting that most of these particles will never be detected.  In order to construct a normalized one-particle distribution, we define
\begin{equation}
\tilde{P}\left(t\right) = P\left(t\right) + \left(1 - C_{\rm{trans}}\right) \delta\left(t - T\right) ,
\end{equation}
where $T$ is larger than any time of interest (we take the $T \rightarrow \infty$ limit, to account for reflected particles). We now write
\begin{equation}
\begin{array}{ll}
P_{\rm{1st}} \left(t\right)&=  \int_{0}^{\infty}dt_{1} ... \int_{0}^{\infty}dt_{N} \tilde{P}\left(t_{1}\right) ... \tilde{P}\left(t_{N}\right) \delta\left(t - \min\left(t_{j}\right)\right) \\[1pt]
&= N! \int_{0}^{\infty}dt_{1} ... \int_{t_{N-1}}^{\infty}dt_{N} \tilde{P}\left(t_{1}\right) ... \tilde{P}\left(t_{N}\right) \delta\left(t - t_{1}\right) \\[1pt]
&= N \tilde{P}\left(t\right) \left(N-1\right)!\!\! \int_{t}^{\infty}dt_{2} ...\!\! \int_{t_{N-1}}^{\infty}dt_{N} \tilde{P}\left(t_{2}\right) ... \tilde{P}\left(t_{N}\right) \\[1pt]
&= N P\left(t\right) \tilde{C}^{N-1}\left(t\right),
\end{array}
\label{P1st}
\end{equation}%
where
\begin{equation}
\tilde{C}\left(t\right) = \int_{t}^{\infty}dt' \tilde{P}\left(t'\right)
\label{notC}
\end{equation}
is the probability of not detecting a particle before time = $t$.

Equation \ref{P1st} is simply interpreted.  The probability of a first click in the time interval $t$ to $t+dt$ is
\begin{equation}
P_{1st}\left(t\right) dt = N P\left(t\right) dt \tilde{C}^{N-1}\left(t\right);
\end{equation}
i.e., the probability that particle 1 is detected in the time interval multiplied by the probability that no other particles have yet been detected.  The $N$ factor arises because any of the $N$ particles could have been detected.  In the context of Brownian-particle first-passage time distributions, this formula appears in \cite{13AsfawAlvarezLacalleShiferaw} (Eq. 15 on p. 6).

Properties of the first-click distribution, derived in the Appendix, are listed as follows.

(1) The first-click distribution integrates to
\begin{equation}
\int_{0}^{\infty}dt P_{1st}\left(t\right)=1-\exp\left(-NC_{\rm{trans}}\right).
\end{equation}
If the number of particles is much greater than $C_{\rm{trans}}^{-1}$, a first click occurs with virtual certainty.

Other properties of the first-click distribution require an explicit model for the $P\left(t\right)$ distribution.  For this purpose, we use the frozen SDA of Eq. (\ref{Psd0}). This approximation is ultimately not good enough to assess whether tunneling electrons can ever have a mean first-click time earlier than for the same number of photons.  But it is sufficient to derive these properties.

(2) If $N \gg 1/C_{\rm{trans}}$, the mean first-click time is given by
\begin{equation}
t_{\rm{1st}}=t_{\rm{mp}}-\delta t \left(\ln\left(\frac{N C_{\rm{trans}}}{2 \pi^{1/2}})\right)\right)^{1/2},
\label{t1st}
\end{equation}
where $t_{\rm{mp}}$ is, as above, the post-selected mean transmission time and
\begin{equation}
\delta t=\frac{\Gamma^{1/2}\left|\Delta^{\sharp }_{\rm{mp}}\right|}{v^{\sharp }_{\rm{mp}}}.
\label{width}
\end{equation}
is the width of $P_{\rm{sd0}}(t)$.  Equation (\ref{width}) shows how the use of many particles advances the detection time.  Since this advancement depends on $N$ only via $N C_{\rm{trans}}$, we see how small transmission probability delays the advancement.

(3) If $N \gg 1/C_{\rm{trans}}$, the width of the first-click distribution is given by
\begin{equation}
\delta t_{\rm{1st}}=\frac{\delta t}{2 \left(e\ln\left(\frac{N C_{\rm{trans}}}{2 \pi^{1/2}}\right)\right)^{1/2}}
\label{width1st}
\end{equation}
This shows the narrowing of the width of detection times resulting from the use of many particles.  Related to this property is the observation that the first-click distribution drops very sharply at times just beyond $t_{1st}$.  This is key to signal transmission, since if no particle is detected by $t_{\rm{1st}}+\delta t_{\rm{1st}}/4$, the signal can almost certainly be interpreted as a binary 0.  We choose one quarter width here to account for the asymmetry of the first-click distribution---it is narrower on the long time side.

(4) Our principal interest is whether the first-click distribution for electrons can peak earlier than that for the same number of free photons---the luminal benchmark. Because the number of particles is typically very large, the first-click distribution peaks when the cumulative probability distribution, $C(t)=1-\tilde{C}(t)$, reaches $1/N$.  This is the time by which observation of at least one particle is likely.  Beyond this time, the high power of the complementary cumulative distribution, $\tilde{C}\left(t\right)$, in Eq. (\ref{P1st}) makes the first-click distribution decay rapidly.  These facts lead to the following observation. If one distribution, for instance the distribution for photons in a vacuum, is larger than another, for instance the distribution for tunneling electrons, then the cumulative probability of the larger distribution will reach $1/N$ first.  The question of which first-click distribution peaks first is equivalent to the question of whether the electron distribution $P(t)$ exceeds the corresponding photon distribution at early times.

If we can show that transmission time distributions for tunneling electrons are smaller than those for photons for all times up to $t_{\rm{mp}}$, then the MacColl-Hartman effect cannot be used to transmit signals faster than is possible using light.  This is what has been found so far, numerically.  Since $P_{\rm{sd0}}(t)$ is analytic, and generally well approximates $P(t)$, it is a natural starting point to support this observation.  However, $P_{\rm{sd0}}(t)$ can exceed $P_{\gamma}(t)$.  To see this, note that because the rate of wave packet broadening is vastly reduced in the relativistic case, the width of the electron and photon distributions are very nearly the same in all computations.  The main differences between them are the reduced normalization of the electron distribution and its earlier postselected mean time.  As $P_{\rm{sd0}}(t)$ and $P_{\gamma}(t)$ are both Gaussians, they are both upside-down parabolas on a logarithmic scale.  The electron's parabola is shifted to the ``right'' and ``down'' relative to the photon's.  If this depiction were truly accurate, then there would be an early time when the electron parabola crosses the photon parabola [see Eq. (\ref{t1st})].

For example, suppose that $\delta t$ is the same for the tunneling electron and the photon---this is approximately true.  The difference between photon and electron mean first-click times is then given by
\begin{multline}
t_{\rm{1st},\gamma}-t_{\rm{1st}}\! =
t_{\rm{mp},\gamma}-t_{\rm{mp}}
-\\ \delta t \left(\sqrt{\ln \left(\frac{N}{2\pi^{1/2}}\right)} - \sqrt{\ln \left(
\frac{N C_{\rm{trans}}}{2 \pi^{1/2}}\right)}\right) \label{t1stDiff0}
\end{multline}
The first difference on the right of Eq. \ref{t1stDiff0} is positive---the MacColl-Hartman effect---while the second difference (in parentheses) is negative.  Because increasing $N$ increases the magnitude of the second term without affecting the first, a sufficiently large $N$ will make the mean first-click time of the electrons earlier than that of the photons.  

As this effect is not seen in accurate numerical calculations of $P(t)$, the $P_{\rm{sd0}}(t)$ model cannot be sufficiently accurate to answer the superluminal signaling question.  However, if we perform a more accurate SDA separately for every $t$, as opposed to expanding about $t_{\rm{mp}}$, and denote this new distribution as  $P_{\rm{sd}}(t)$, we find that this quantity never exceeds $P_{\gamma}(t)$.

\section{Numerical Results}

\begin{figure}
\includegraphics[width=0.5\textwidth]{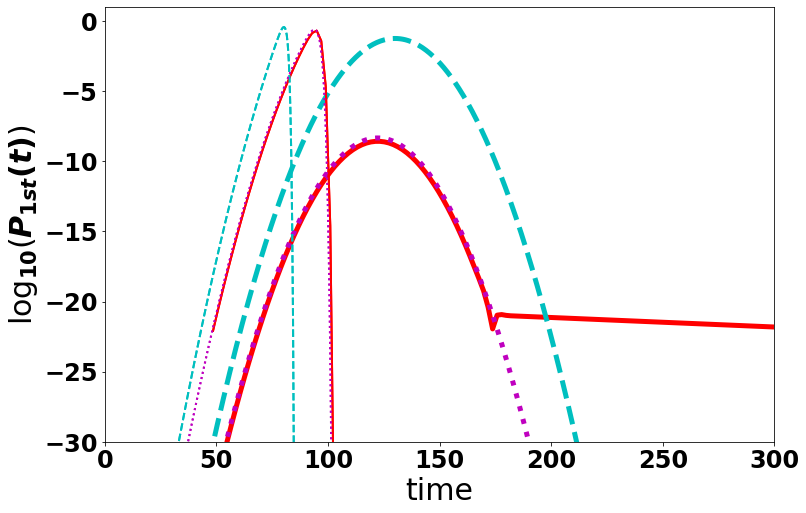}
\includegraphics[width=0.5\textwidth]{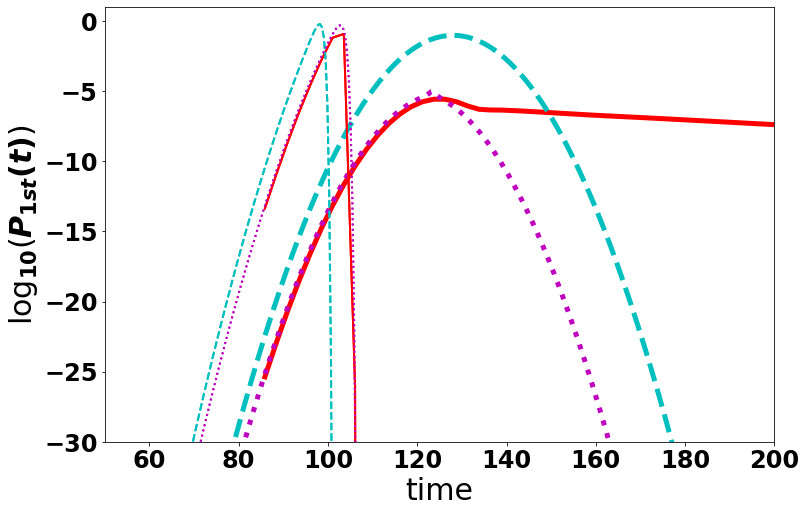}
\caption{Electron transmission time distribution (thick solid lines) and associated first-click distribution (thin solid lines) for $N=10^{12}$.  The dotted lines are the corresponding SDAs.  Also shown are corresponding transmission time and first-click distributions for photons propagating in a vacuum (dashed thick and thin lines, respectively).  In both panels, the initial wave packet is 120 $\lambdabar$ (1 $\lambdabar = 386$ fm: the reduced Compton wavelength of an electron) to the right of the barrier with mean velocity $0.99 c$. The unit of time is $\lambdabar/ c = 1.29$ zs.  In the top panel, the initial wavepacket width and barrier width are both 10 $\lambdabar$.  The barrier height is 7.5 $m c^{2}$.  The accurate numerical transmission time distribution is computed using a grid of $10^{5}$ points.  It is cut off at its short time accuracy limit, $10^{-36}$.  In the bottom panel, the initial wave packet and barrier widths are 6 and 8 $\lambdabar$, respectively.  The barrier height is 6.52 $m c^{2}$.  The grid had $10^{6}$ points, giving a short time accuracy limit of $10^{-27}$.}
\label{fig:1stclickdistributions}
\end{figure}

The first-click (for $N=10^{12}$ particles) and corresponding $P(t)$ distributions are shown in Fig.~\ref{fig:1stclickdistributions}.  Corresponding distributions for photons in a vacuum are also shown.  We find that the suppression of the tunneling time distribution by the tunneling probability manifests as a delay in the advancement of the first-click time until $N$ exceeds $C_{\rm{trans}}^{-1}$.  In particular, though the transmission time distribution peaks earlier than the equivalent distribution for photons, the first-click distribution for photons always peaks earlier. Also shown (dotted lines) are the tunneling time distributions (and associated first-click distributions) computed using the steepest descent method at every time, $P_{\rm{sd}}(t)$.

The initial wave packet widths for the top and bottom panels of Fig.~\ref{fig:1stclickdistributions} are approximately 11 and 7 de Broglie wavelengths, respectively.  The bottom panel is close to a best-case scenario: it is narrow in position space to reduce the width of $P(t)$ and enhance the impact of the MacColl-Hartman time advance. Any alterations would diminish the MacColl-Hartman effect: a higher barrier energy or an increased barrier width would decrease the tunneling probability, a decreased barrier width would decrease the MacColl-Hartman time advance, a lower barrier energy would produce mostly above-barrier transmission, and a wave packet narrower in position space would be so broad in momentum space that the above-barrier and Klein zone contributions would dominate. In all cases, the impact of the MacColl-Hartman effect is reduced. 

The observed distributions are composed of a Gaussian peak at short times plus a slowly decaying exponential tail at longer times. The tail arises from the above-barrier portion of the initial wave packet, which is not modeled in our current SDA, as we only considered the below-barrier contribution to it. There is, in fact, a series of saddle points, only the first of which corresponds to the tunneling contribution below the barrier. The others all correspond to above-barrier contributions with increasing numbers of oscillations in the imaginary direction in space.  Each saddle point would contribute a Gaussian in time to the overall distribution, but with amplitude exponentially small in the number of imaginary space oscillations.  When combined, they would produce an exponentially decaying long-time tail with negligible contribution at short times.

To see why $P_{\rm{sd}}(t)$ never crosses $P_{\gamma}(t)$, while $P_{\rm{sd0}}(t)$ does, suppose we compute an SDA at some time, $t_1$, other than the most probable time.  Just as $P_{\rm{sd0}}(t)$ is obtained by expanding $\ln\left(P_{\rm{sd}}(t)\right)$ to second order around $t_{\rm{mp}}$, a model $P_{\rm{sd1}}(t)$ can be obtained by expanding instead about $t_1$.  $\ln P_{\rm{sd1}}(t)$ is an upside-down parabola, just like $\ln P_{\rm{sd0}}(t)$.  The difference is that the parameters of the parabola are determined by the SDA at a different value of the physical time.  Equation (\ref{psharp}) shows that the physical time enters into the calculation only through its difference with ${\rm{Re}} \tau(p^{\sharp})$.  As ${\rm{Re}} \tau(p^{\sharp})$ varies, the effective center of the upside-down parabola varies.  In particular, at early times ${\rm{Re}} \tau(p^{\sharp})$ shifts from its MacColl-Hartman advanced value to a later time corresponding to subluminal effective speed.

\begin{figure}[t]
\includegraphics[width=1.0\linewidth]{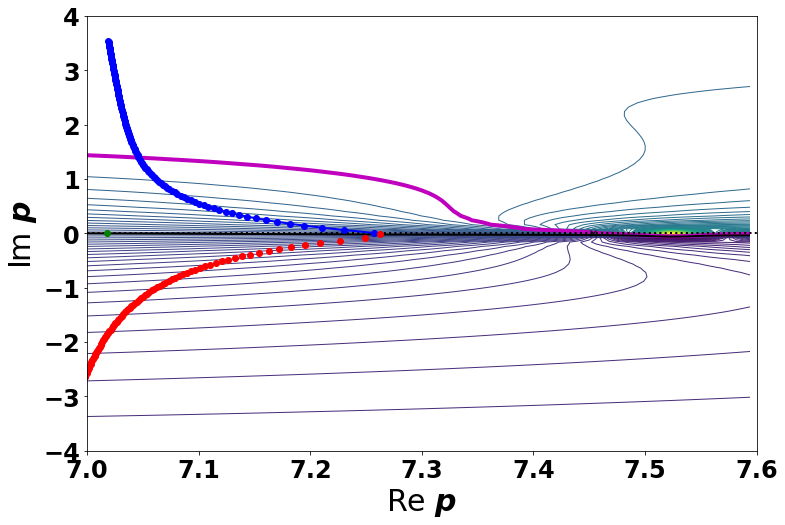}
\caption{Contour lines of ${\rm{Re}} \tau(p)$ in the complex $p$ plane for the wave packet and barrier of the bottom panel of Fig. 1.  The thick contour corresponds to ${\rm{Re}} \tau(p)=128$, which is how long it would take a photon traveling in a vacuum to reach the observation point.  Also shown is the path (large dots) taken by $p^{\sharp}$.  It starts ($t=0$) in the upper left corner, swoops down and to the right to the real axis when $t=t_{\rm{mp}}$, then swoops down and back to the left at longer times.  The dots are plotted every 10 $\lambdabar/c$.}
\label{fig:taucontrs}
\end{figure}

\begin{figure}[t]
\includegraphics[width=1.0\linewidth]{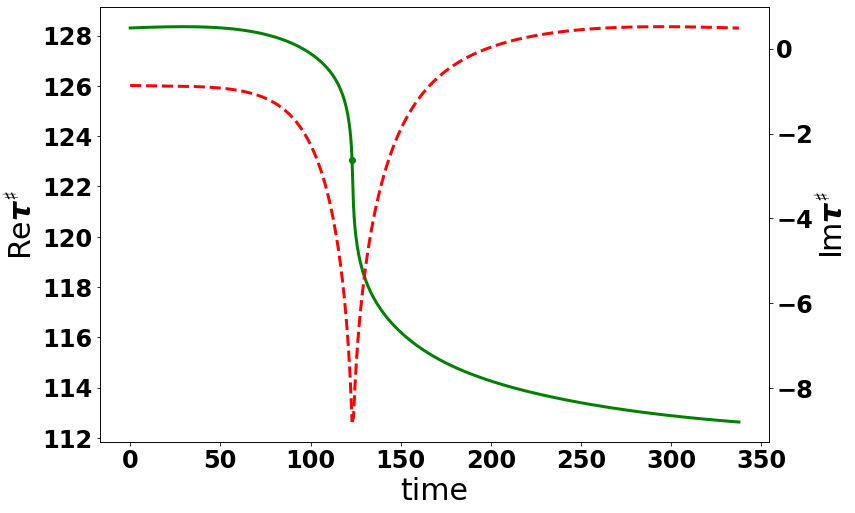}
\caption{Real (solid line) and imaginary (dashed line) parts of $\tau^{\sharp}$ in the complex-$p$ plane as functions of the physical time, $t$, for the steepest-descent calculations shown in the bottom panel of Fig.~\ref{fig:1stclickdistributions}. The dot shows ${\rm{Re}} \tau^{\sharp}$ at the most probable time. The real and imaginary parts of $\tau^{\sharp}$ correspond respectively to the effective Wigner phase times and Pollak-Miller imaginary times.}
\label{fig:tausharp}
\end{figure}

To understand the early-time behavior of ${\rm{Re}} \tau^{\sharp}$, we first study $p^{\sharp}$ at these times.  From Eq. (\ref{psharp}) we see that, as $t$ moves away from $t_{\rm{mp}}$, only the imaginary part of $p^{\sharp}$ varies.  For early times, the imaginary part of $p^{\sharp}$ is positive.  Figure \ref{fig:taucontrs} shows the path of $p^{\sharp}$ in the complex momentum plane and the contour lines of ${\rm{Re}} \tau(p)$, while Fig.~\ref{fig:tausharp} focuses on just $\tau^{\sharp}$ as it varies with physical time $t$.  

In Fig.~\ref{fig:taucontrs}, on the real $p$ axis below $p_{\rm{top}}$ (the momentum of the barrier height), ${\rm{Re}} \tau(p)$ is almost constant---equal to its MacColl-Hartman advanced value.  $\tau(p)$ varies in the negative imaginary direction when $p$ increases on the real axis towards $p_{\rm{top}}$.  If $p$ varies in the positive imaginary direction (corresponding to the case of early times), then Fig.~\ref{fig:taucontrs} shows that ${\rm{Re}}\tau(p)$ increases. 

As $t$ gets closer to zero, $p^{\sharp}$ varies more slowly. The associated ${\rm{Re}}\left(\tau(p^{\sharp})\right)$ values are even less variable---in Fig.~\ref{fig:taucontrs}, the contours of ${\rm{Re}}\left(\tau(p^{\sharp})\right)$ become widely spaced (as would the contours of ${\rm{Im}}\left(\tau(p^{\sharp})\right)$ if plotted) and, in Fig.~\ref{fig:tausharp}, the plots of ${\rm{Re}}\left(\tau^{\sharp}\right)$ and ${\rm{Im}}\left(\tau^{\sharp}\right)$ as functions of physical time become flat. Indeed, Fig.~\ref{fig:tausharp} shows that, as $t \rightarrow 0$, ${\rm{Re}}\left(\tau(p^{\sharp})\right)$ approaches the time a free particle traveling $0.998c$ would take, and this time also matches the time one would expect given momentum filtering of the initial $0.99c$ velocity wave packet.

The contours of ${\rm{Im}}\left(\tau(p)\right)$ (not shown in Fig.~\ref{fig:taucontrs}) feature a sharp trough along the real axis.  What appears to be a cusp in Fig.~\ref{fig:tausharp} for ${\rm{Im}}\left(\tau(p^{\sharp})\right)$ at $t_{\rm{mp}}$ is actually just a very sharp minimum.  The sharpness results from the rapid variation in $p^{\sharp}$ near the most probable time, and the sharpness of the ${\rm{Im}}\left(\tau(p)\right)$ real axis trough in the $p$ plane.

As an aside, we also note that although this work is focused on the fading of the MacColl-Hartman advancement at early times, we also see \textit{enhancement} of the advancement at later times in Fig.~\ref{fig:tausharp}. The net effect of this is that the tunneled wave packet is slightly ``compressed'' in comparison to a free particle wave packet.

\section{Conclusions}

Other authors have advanced many arguments to explain away the supposed superluminal transmission in quantum tunneling, such as subtle ideas about the tunneling process reshaping the wave packet \cite{97ChiaoSteinberg}. Some authors have also argued that the ``front'' of the wave packet \cite{03Winful, 03ButtikerWashburn} is the source of the tunneled particles to avoid speeds faster than light. In a sense, this work corroborates those assertions, as the ``first-click distribution'' is essentially a way of characterizing that ``front.'' However, the arguments presented here do not make claims about which particles in the distribution tunnel, but only what can be said about the statistics of their time of arrival, showing that they can never ``out-compete'' free-traveling photons.

In our previous works, we clarified the relevance of different definitions to genuine, observable tunneling times and showed which statements can be definitively made about wave-packet reshaping in tunneling. Here, we expand that argument, elaborating on the ``signaling'' aspect of the debate. By making explicit how tunneling particles could be used to send a signal as fast as possible, we find the true requirement for superluminal signaling via the MacColl-Hartman effect.  Specifically, the (unnormalized) particle transmission time distribution must exceed that for photons in a vacuum for some early time (well ahead of the peak time). Having not seen this for any of our accurate computations, we explained the observation by showing that the MacColl-Hartman time advance fades away when considering the short times relevant to signaling via large numbers of particles.  The MacColl-Hartman time advance manifests in the phase time computed using the SDA, when time is near the peak time. At early times, the SDA gives an effective phase time without a time advance.

Results presented here are for tunneling in one dimension. Such a model arises naturally for an electron in three-dimensional space directly impacting a barrier varying in only one direction.  We do not expect that adding $x$ and $y$ dependence to the potential would impact the conclusions of this paper.  While studies in multiple dimensions are still very much of interest \cite{98BracherBeckerGurvitz}, it would be much more difficult to achieve the high level of accuracy required here to investigate deep tunneling at early times.  Also note that, in any case, multidimensional tunneling typically reduces to the computation of instantons, which are one-dimensional curvilinear pathways through barriers \cite{23NandiLaudeKhire}.

We end by noting that, although the nonrelativistic case was not studied in this work, we would anticipate similar conclusions to be reached were one to perform equivalent calculations there---the most meaningful physical difference between the two cases is that in the non-relativistic case, the momentum dispersion effects are much larger, which serves to further suppress the MacColl-Hartman effect. This would further reduce the capacity of a signal composed of tunneled particles to arrive ahead of a signal composed of free-traveling particles.

\begin{acknowledgements}
We wish to thank E. Pollak and S. Miret-Art\'es for many useful discussions. R.S.D. acknowledges support from the Professional Development Allowance of McMaster University. T.R. acknowledges support from the Israel Science Foundation (project code 408/19), and from a Senior Postdoctoral Fellowship of the Weizmann Institute of Science’s Feinberg Graduate School.
\end{acknowledgements}

\section*{Appendix: Properties of the First-Click Distribution}

From Eqs. (\ref{P1st}) and (\ref{notC}),
\begin{equation}
\begin{array}{ll}
P_{\rm{1st}}\left(t\right)&= NP(t)\left(1-\frac{NC(t)}{N}\right)^{N-1} \\[1pt]
&= \frac{P(t)}{C(t)}NC(t)\exp\left(-NC(t)\right) \\[1pt]
&=\frac{d}{dt}{\rm{ln}}\left(C(t)\right) f\left(NC(t)\right)
\end{array}
\label{P1st_largeN}
\end{equation}
where $f(x)=x\exp(-x)$. The function, $f(x)$, has a maximum at $x=1$, with associated width
\begin{equation}
\delta x = \left(-\frac{d^{2}}{dx^{2}}{\rm{ln}}\left(f(x)\right)|_{x=1}\right)^{1/2}=e^{1/2}.
\end{equation}
If $N \gg C_{\rm{trans}}^{-1}$, $f(NC(t))$ has a sharp maximum when $C(t)=1/N$; i.e.,
\begin{equation}
\int_{0}^{t_{\rm{1st}}} dt_{1} P(t_{1}) =\frac{1}{N}.
\label{t1st_implicit}
\end{equation}

Since the first factor in the expression for $P_{\rm{1st}}(t)$, in Eq. \ref{P1st_largeN}, is slowly varying compared to the second term, the peak in the first arrival time distribution occurs at the time of the peak in $f$---i.e., when the cumulative probability distribution reaches 1/N.  The width associated with this peak is determined by
\[
NC'(t)\delta t_{\rm{1st}}=\delta x
\]
or
\begin{equation}
\delta t_{\rm{1st}}=\frac{1}{N} \frac{1}{e^{1/2}P(t_{\rm{1st}})}.
\label{deltat1st}
\end{equation}

To get the total probability of a first click we integrate $P_{\rm{1st}}(t)$ over all time.  This simplifies if Eq. (\ref{P1st_largeN}) is written in the following equivalent form,
\begin{equation}
P_{\rm{1st}}\left(t\right) = f\left(ab(t)\right) \frac{d}{dt}{\rm{ln}}\left(b(t)\right)
\end{equation}
where $a=NC_{\rm{trans}}$ and $b(t)=C(t)/C_{\rm{trans}}$.  Using this expression we get
\begin{equation}
\begin{array}{ll}
\int_{0}^{\infty} dt P_{\rm{1st}}\left(t\right)= \int_{0}^{\infty}dt f\left(ab(t)\right) \frac{d}{dt}{\rm{ln}}\left(b(t)\right) \\
= \left[ f\left(ab(t)\right){\rm{ln}} \left(b(t)\right)\right]_{0}^{\infty}-\int_{0}^{\infty} dt \left(\frac{d}{dt}f(ab(t))\right){\rm{ln}}\left(b(t)\right)\\
=-a\int_{0}^{1}db(1-ab){\rm{exp}}(-ab){\rm{ln}}(b)\\
=\int_{0}^{a}dx(x-1){\rm{exp}}(-x)\left({\rm{ln}}(x)-{\rm{ln}}(a)\right)\\
=\left[-\left(x{\rm{ln}}(x)+1\right){\rm{exp}}(-x)\right]_{0}^{a}-{\rm{ln}}(a)[-x {\rm{exp}}(-x)]_{0}^{a}\\
=1-{\rm{exp}}\left(-a\right)=1-{\rm{exp}}\left(-NC_{\rm{trans}}\right).
\end{array}
\label{eq:blah}
\end{equation}
Here, we see that, for $N$ much larger than $1/C_{\rm{trans}}$, a first click event becomes virtually certain.

To further characterize the first click distribution we need an explicit form for the tunneling time distribution.  This is provided by the steepest descent model, $P_{\rm{sd0}}(t)$ as a Gaussian of the form,
\begin{equation}
P_{\rm{sd0}}(t)=\frac{C_{\rm{trans}}}{\pi^{1/2}\delta t}\, {\rm{exp}}\left(-\left(\frac{t-t_{\rm{mp}}}{\delta t}\right)^{2}\right).
\end{equation}
In this case, defining $t_{\rm{1dif}}=\left(t_{\rm{1st}}-t_{\rm{mp}}\right)/\delta t$, the mean first click time is determined by
\begin{equation}
\frac{1}{N}=\frac{C_{\rm{trans}}}{\pi^{1/2}}\int_{0}^{t_{\rm{1dif}}}dx{\rm{exp}}(-x^{2})
=\frac{C_{\rm{trans}}}{2}\,{\rm{erf}}\left(t_{\rm{1dif}}\right).
\end{equation}
If $N \gg 1/C_{\rm{trans}}$, we can use the asymptotic form of the error function to get
\begin{equation}
\frac{1}{N}=\frac{C_{\rm{trans}}}{2\pi^{1/2}}\frac{{\rm{exp}}\left(t_{\rm{1dif}}\right)}{t_{\rm{1dif}}}.
\end{equation}
This leads to
\begin{equation}
t_{\rm{1st}}=t_{\rm{mp}}-x\delta t
\end{equation}
where $x$ is the solution to the nonlinear equation,
\begin{equation}
x=\frac{a}{2\pi^{1/2}}{\rm{exp}}(-x^{2})
\end{equation}
or
\begin{equation}
x=\left({\rm{ln}}\left(\frac{a}{2\pi^{1/2}}\right)-{\rm{ln}}(x)\right)^{1/2}.
\end{equation}
The latter equation can be solved by iteration, if $a > 0.519$.  To leading order (for large $a$),
\begin{equation}
t_{\rm{1st}}=t_{\rm{mp}}-\delta t\left({\rm{ln}}\left(\frac{NC_{\rm{trans}}}{2\pi^{1/2}}\right)\right)^{1/2}
\end{equation}
So, the mean of the first click distribution is advanced to earlier than $t_{\rm{mp}}$ in proportion to $\left({\rm{ln}}(NC_{\rm{trans}})\right)^{1/2}$, when $N$ is very large.

The width of the first click distribution is also made explicit by the frozen steepest descent model.  From Eq. \ref{deltat1st} and the above expression for $P_{\rm{sd0}}(t)$, we get
\begin{equation}
\delta t_{\rm{1st}} =\frac{\delta t}{2e^{1/2}\left({\rm{ln}}\left(\frac{NC_{\rm{trans}}}{2\pi^{1/2}}\right)\right)^{1/2}}
\end{equation}
when $N$ is large.  We see that the first click distribution gradually becomes narrower, with increasing $N$.\\


\begin{thebibliography}{66}%
\makeatletter
\providecommand \@ifxundefined [1]{%
 \@ifx{#1\undefined}
}%
\providecommand \@ifnum [1]{%
 \ifnum #1\expandafter \@firstoftwo
 \else \expandafter \@secondoftwo
 \fi
}%
\providecommand \@ifx [1]{%
 \ifx #1\expandafter \@firstoftwo
 \else \expandafter \@secondoftwo
 \fi
}%
\providecommand \natexlab [1]{#1}%
\providecommand \enquote  [1]{``#1''}%
\providecommand \bibnamefont  [1]{#1}%
\providecommand \bibfnamefont [1]{#1}%
\providecommand \citenamefont [1]{#1}%
\providecommand \href@noop [0]{\@secondoftwo}%
\providecommand \href [0]{\begingroup \@sanitize@url \@href}%
\providecommand \@href[1]{\@@startlink{#1}\@@href}%
\providecommand \@@href[1]{\endgroup#1\@@endlink}%
\providecommand \@sanitize@url [0]{\catcode `\\12\catcode `\$12\catcode
  `\&12\catcode `\#12\catcode `\^12\catcode `\_12\catcode `\%12\relax}%
\providecommand \@@startlink[1]{}%
\providecommand \@@endlink[0]{}%
\providecommand \url  [0]{\begingroup\@sanitize@url \@url }%
\providecommand \@url [1]{\endgroup\@href {#1}{\urlprefix }}%
\providecommand \urlprefix  [0]{URL }%
\providecommand \Eprint [0]{\href }%
\providecommand \doibase [0]{https://doi.org/}%
\providecommand \selectlanguage [0]{\@gobble}%
\providecommand \bibinfo  [0]{\@secondoftwo}%
\providecommand \bibfield  [0]{\@secondoftwo}%
\providecommand \translation [1]{[#1]}%
\providecommand \BibitemOpen [0]{}%
\providecommand \bibitemStop [0]{}%
\providecommand \bibitemNoStop [0]{.\EOS\space}%
\providecommand \EOS [0]{\spacefactor3000\relax}%
\providecommand \BibitemShut  [1]{\csname bibitem#1\endcsname}%
\let\auto@bib@innerbib\@empty
\bibitem [{\citenamefont {Pollak}\ and\ \citenamefont
  {Miller}(1984)}]{84PollakMiller}%
  \BibitemOpen
  \bibfield  {author} {\bibinfo {author} {\bibfnamefont {E.}~\bibnamefont
  {Pollak}}\ and\ \bibinfo {author} {\bibfnamefont {W.~H.}\ \bibnamefont
  {Miller}},\ }\bibfield  {title} {\bibinfo {title} {New Physical
  Interpretation for Time in Scattering Theory},\ }\href@noop {} {\bibfield
  {journal} {\bibinfo  {journal} {Phys. Rev. Lett.}\ }\textbf {\bibinfo
  {volume} {53}},\ \bibinfo {pages} {115} (\bibinfo {year} {1984})}\BibitemShut
  {NoStop}%
\bibitem [{\citenamefont {Hauge}\ and\ \citenamefont
  {St{\o}vneng}(1989)}]{89HaugeStovneng}%
  \BibitemOpen
  \bibfield  {author} {\bibinfo {author} {\bibfnamefont {E.}~\bibnamefont
  {Hauge}}\ and\ \bibinfo {author} {\bibfnamefont {J.}~\bibnamefont
  {St{\o}vneng}},\ }\bibfield  {title} {\bibinfo {title} {Tunneling times: a
  critical review},\ }\href@noop {} {\bibfield  {journal} {\bibinfo  {journal}
  {Rev. Mod. Phys.}\ }\textbf {\bibinfo {volume} {61}},\ \bibinfo {pages} {917}
  (\bibinfo {year} {1989})}\BibitemShut {NoStop}%
\bibitem [{\citenamefont {Low}\ and\ \citenamefont {Mende}(1991)}]{91LowMende}%
  \BibitemOpen
  \bibfield  {author} {\bibinfo {author} {\bibfnamefont {F.~E.}\ \bibnamefont
  {Low}}\ and\ \bibinfo {author} {\bibfnamefont {P.~F.}\ \bibnamefont
  {Mende}},\ }\bibfield  {title} {\bibinfo {title} {A note on the tunneling
  time problem},\ }\href@noop {} {\bibfield  {journal} {\bibinfo  {journal}
  {Ann. Phys.-New York}\ }\textbf {\bibinfo {volume} {210}},\ \bibinfo {pages}
  {380} (\bibinfo {year} {1991})}\BibitemShut {NoStop}%
\bibitem [{\citenamefont {Muga}\ \emph {et~al.}(1992)\citenamefont {Muga},
  \citenamefont {Brouard},\ and\ \citenamefont {Sala}}]{92MugaBrouard}%
  \BibitemOpen
  \bibfield  {author} {\bibinfo {author} {\bibfnamefont {J.}~\bibnamefont
  {Muga}}, \bibinfo {author} {\bibfnamefont {S.}~\bibnamefont {Brouard}},\ and\
  \bibinfo {author} {\bibfnamefont {R.}~\bibnamefont {Sala}},\ }\bibfield
  {title} {\bibinfo {title} {Transmission and reflection tunneling times},\
  }\href@noop {} {\bibfield  {journal} {\bibinfo  {journal} {Phys. Lett.
  A}\ }\textbf {\bibinfo {volume} {167}},\ \bibinfo {pages} {24} (\bibinfo
  {year} {1992})}\BibitemShut {NoStop}%
\bibitem [{\citenamefont {Leavens}\ and\ \citenamefont
  {Aers}(1993)}]{93Leavens}%
  \BibitemOpen
  \bibfield  {author} {\bibinfo {author} {\bibfnamefont {C.~R.}\ \bibnamefont
  {Leavens}}\ and\ \bibinfo {author} {\bibfnamefont {G.~C.}\ \bibnamefont
  {Aers}},\ }\bibfield  {title} {\bibinfo {title} {Bohm trajectories and the
  tunneling time problem},\ }in\ \href@noop {} {\emph {\bibinfo {booktitle}
  {Scanning Tunneling Microscopy III}}}\ (\bibinfo  {publisher} {Springer},\
  \bibinfo {year} {1993})\ pp.\ \bibinfo {pages} {105--140}\BibitemShut
  {NoStop}%
\bibitem [{\citenamefont {Brouard}\ \emph {et~al.}(1994)\citenamefont
  {Brouard}, \citenamefont {Sala},\ and\ \citenamefont {Muga}}]{94BrouardSala}%
  \BibitemOpen
  \bibfield  {author} {\bibinfo {author} {\bibfnamefont {S.}~\bibnamefont
  {Brouard}}, \bibinfo {author} {\bibfnamefont {R.}~\bibnamefont {Sala}},\ and\
  \bibinfo {author} {\bibfnamefont {J.~G.}~\bibnamefont {Muga}},\ }\bibfield
  {title} {\bibinfo {title} {Systematic approach to define and classify quantum
  transmission and reflection times},\ }\href@noop {} {\bibfield  {journal}
  {\bibinfo  {journal} {Phys. Rev. A.}\ }\textbf {\bibinfo {volume} {49}},\
  \bibinfo {pages} {4312} (\bibinfo {year} {1994})}\BibitemShut {NoStop}%
\bibitem [{\citenamefont {Steinberg}(1995{\natexlab{a}})}]{95Steinberga}%
  \BibitemOpen
  \bibfield  {author} {\bibinfo {author} {\bibfnamefont {A.~M.}\ \bibnamefont
  {Steinberg}},\ }\bibfield  {title} {\bibinfo {title} {How Much Time Does a
  Tunneling Particle Spend in the Barrier Region?},\ }\href@noop {} {\bibfield
  {journal} {\bibinfo  {journal} {Phys. Rev. Lett.}\ }\textbf {\bibinfo
  {volume} {74}},\ \bibinfo {pages} {2405} (\bibinfo {year}
  {1995}{\natexlab{a}})}\BibitemShut {NoStop}%
\bibitem [{\citenamefont {Steinberg}(1995{\natexlab{b}})}]{95Steinbergb}%
  \BibitemOpen
  \bibfield  {author} {\bibinfo {author} {\bibfnamefont {A.~M.}\ \bibnamefont
  {Steinberg}},\ }\bibfield  {title} {\bibinfo {title} {Conditional
  probabilities in quantum theory and the tunneling-time controversy},\
  }\href@noop {} {\bibfield  {journal} {\bibinfo  {journal} {Phys. Rev. A.}\
  }\textbf {\bibinfo {volume} {52}},\ \bibinfo {pages} {32} (\bibinfo {year}
  {1995}{\natexlab{b}})}\BibitemShut {NoStop}%
\bibitem [{\citenamefont {Hauge}(1997)}]{97Hauge}%
  \BibitemOpen
  \bibfield  {author} {\bibinfo {author} {\bibfnamefont {E.~H.}\ \bibnamefont
  {Hauge}},\ }\bibfield  {title} {\bibinfo {title} {Can one speak about
  tunneling times in polite society?},\ }in\ \href@noop {} {\emph {\bibinfo
  {booktitle} {Proceedings of the Adriatico Research Conference on Tunneling and Its Implications}}}\
  \bibinfo {editor} {edited by\ \bibinfo
  {editor} {\bibfnamefont {D.}~\bibnamefont {Mugnai}}, \bibinfo {editor}
  {\bibfnamefont {A.}\ \bibnamefont {Ranfagni}},\ and\ \bibinfo {editor}
  {\bibfnamefont {L.}~\bibnamefont {Schulman}}}\
  (\bibinfo  {publisher} {World
  Scientific, Singapore},\ \bibinfo {year} {1997})\ pp.\ \bibinfo {pages}
  {1--17}\BibitemShut {NoStop}%
\bibitem [{\citenamefont {Muga}\ and\ \citenamefont
  {Leavens}(2000)}]{00MugaLeavens}%
  \BibitemOpen
  \bibfield  {author} {\bibinfo {author} {\bibfnamefont {J.~G.}\ \bibnamefont
  {Muga}}\ and\ \bibinfo {author} {\bibfnamefont {C.~R.}\ \bibnamefont
  {Leavens}},\ }\bibfield  {title} {\bibinfo {title} {Arrival time in quantum
  mechanics},\ }\href@noop {} {\bibfield  {journal} {\bibinfo  {journal} {Phys.
  Rep.}\ }\textbf {\bibinfo {volume} {338}},\ \bibinfo {pages} {353} (\bibinfo
  {year} {2000})}\BibitemShut {NoStop}%
\bibitem [{\citenamefont {Ruseckas}(2001)}]{01Ruseckas}%
  \BibitemOpen
  \bibfield  {author} {\bibinfo {author} {\bibfnamefont {J.}~\bibnamefont
  {Ruseckas}},\ }\bibfield  {title} {\bibinfo {title} {Possibility of tunneling
  time determination},\ }\href@noop {} {\bibfield  {journal} {\bibinfo
  {journal} {Phys. Rev. A.}\ }\textbf {\bibinfo {volume} {63}},\ \bibinfo
  {pages} {052107} (\bibinfo {year} {2001})}\BibitemShut {NoStop}%
\bibitem [{\citenamefont {Ruseckas}\ and\ \citenamefont
  {Kaulakys}(2001)}]{01RuseckasKaulakys}%
  \BibitemOpen
  \bibfield  {author} {\bibinfo {author} {\bibfnamefont {J.}~\bibnamefont
  {Ruseckas}}\ and\ \bibinfo {author} {\bibfnamefont {B.}~\bibnamefont
  {Kaulakys}},\ }\bibfield  {title} {\bibinfo {title} {Time problem in quantum
  mechanics and weak measurements},\ }\href@noop {} {\bibfield  {journal}
  {\bibinfo  {journal} {Phys. Lett. A}\ }\textbf {\bibinfo {volume}
  {287}},\ \bibinfo {pages} {297} (\bibinfo {year} {2001})}\BibitemShut
  {NoStop}%
\bibitem [{\citenamefont {Winful}(2003)}]{03Winful}%
  \BibitemOpen
  \bibfield  {author} {\bibinfo {author} {\bibfnamefont {H.~G.}\ \bibnamefont
  {Winful}},\ }\bibfield  {title} {\bibinfo {title} {Delay Time and the Hartman
  Effect in Quantum Tunneling},\ }\href@noop {} {\bibfield  {journal} {\bibinfo
   {journal} {Phys. Rev. Lett.}\ }\textbf {\bibinfo {volume} {91}},\ \bibinfo
  {pages} {260401} (\bibinfo {year} {2003})}\BibitemShut {NoStop}%
\bibitem [{\citenamefont {Muga}(2008)}]{08Muga}%
  \BibitemOpen
  \bibfield  {author} {\bibinfo {author} {\bibfnamefont {J.~G.}\ \bibnamefont
  {Muga}},\ }\bibfield  {title} {\bibinfo {title} {Characteristic times in
  one-dimensional scattering},\ }in\ \href@noop {} {\emph {\bibinfo {booktitle}
  {Time in Quantum Mechanics}}},\ \bibinfo {editor} {edited by\ \bibinfo
  {editor} {\bibfnamefont {J.}~\bibnamefont {Muga}}, \bibinfo {editor}
  {\bibfnamefont {R.~S.}\ \bibnamefont {Mayato}},\ and\ \bibinfo {editor}
  {\bibfnamefont {{\'I}.}~\bibnamefont {Egusquiza}}}\ (\bibinfo  {publisher}
  {Springer Berlin Heidelberg},\ \bibinfo {address} {Berlin, Heidelberg},\
  \bibinfo {year} {2008})\ pp.\ \bibinfo {pages} {31--72}\BibitemShut {NoStop}%
\bibitem [{\citenamefont {Egusquiza}\ \emph {et~al.}(2008)\citenamefont
  {Egusquiza}, \citenamefont {Muga},\ and\ \citenamefont
  {Baute}}]{08EgusquizaMuga}%
  \BibitemOpen
  \bibfield  {author} {\bibinfo {author} {\bibfnamefont {I.~L.}\ \bibnamefont
  {Egusquiza}}, \bibinfo {author} {\bibfnamefont {J.~G.}\ \bibnamefont
  {Muga}},\ and\ \bibinfo {author} {\bibfnamefont {A.~D.}\ \bibnamefont
  {Baute}},\ }\bibfield  {title} {\bibinfo {title} {``standard"
  quantum--mechanical approach to times of arrival},\ }in\ \href@noop {} {\emph
  {\bibinfo {booktitle} {Time in Quantum Mechanics}}}\ (\bibinfo  {publisher}
  {Springer},\ \bibinfo {year} {2008})\ pp.\ \bibinfo {pages}
  {305--332}\BibitemShut {NoStop}%
\bibitem [{\citenamefont {Mayato}\ \emph {et~al.}(2008)\citenamefont {Mayato},
  \citenamefont {Alonso},\ and\ \citenamefont {Egusquiza}}]{08MayatoAlonso}%
  \BibitemOpen
  \bibfield  {author} {\bibinfo {author} {\bibfnamefont {R.~S.}\ \bibnamefont
  {Mayato}}, \bibinfo {author} {\bibfnamefont {D.}~\bibnamefont {Alonso}},\
  and\ \bibinfo {author} {\bibfnamefont {I.~L.}\ \bibnamefont {Egusquiza}},\
  }\bibfield  {title} {\bibinfo {title} {Quantum clocks and stopwatches},\ }in\
  \href@noop {} {\emph {\bibinfo {booktitle} {Time in Quantum Mechanics}}}\
  (\bibinfo  {publisher} {Springer},\ \bibinfo {year} {2008})\ pp.\ \bibinfo
  {pages} {235--278}\BibitemShut {NoStop}%
\bibitem [{\citenamefont {Schulman}(2008)}]{08Schulman}%
  \BibitemOpen
  \bibfield  {author} {\bibinfo {author} {\bibfnamefont {L.~S.}\ \bibnamefont
  {Schulman}},\ }\bibfield  {title} {\bibinfo {title} {Jump time and passage
  time: The duration of a quantum transition},\ }in\ \href@noop {} {\emph
  {\bibinfo {booktitle} {Time in Quantum Mechanics}}}\ (\bibinfo  {publisher}
  {Springer},\ \bibinfo {year} {2008})\ pp.\ \bibinfo {pages}
  {107--128}\BibitemShut {NoStop}%
\bibitem [{\citenamefont {Sokolovski}(2008)}]{08SokolovskiMuga}%
  \BibitemOpen
  \bibfield  {author} {\bibinfo {author} {\bibfnamefont {D.}~\bibnamefont
  {Sokolovski}},\ }\bibfield  {title} {\bibinfo {title} {Quantum traversal
  time, path integrals and “superluminal” tunnelling},\ }in\ \href@noop {}
  {\emph {\bibinfo {booktitle} {Time in Quantum Mechanics}}}\ (\bibinfo
  {publisher} {Springer},\ \bibinfo {year} {2008})\ pp.\ \bibinfo {pages}
  {195--233}\BibitemShut {NoStop}%
\bibitem [{\citenamefont {Galapon}(2012)}]{12Galapon}%
  \BibitemOpen
  \bibfield  {author} {\bibinfo {author} {\bibfnamefont {E.~A.}\ \bibnamefont
  {Galapon}},\ }\bibfield  {title} {\bibinfo {title} {Only Above Barrier Energy
  Components Contribute to Barrier Traversal Time},\ }\href@noop {} {\bibfield
  {journal} {\bibinfo  {journal} {Phys. Rev. Lett.}\ }\textbf {\bibinfo
  {volume} {108}},\ \bibinfo {pages} {170402} (\bibinfo {year}
  {2012})}\BibitemShut {NoStop}%
\bibitem [{\citenamefont {McDonald}\ \emph {et~al.}(2015)\citenamefont
  {McDonald}, \citenamefont {Orlando}, \citenamefont {Vampa},\ and\
  \citenamefont {Brabec}}]{15McDonaldOrlando}%
  \BibitemOpen
  \bibfield  {author} {\bibinfo {author} {\bibfnamefont {C.}~\bibnamefont
  {McDonald}}, \bibinfo {author} {\bibfnamefont {G.}~\bibnamefont {Orlando}},
  \bibinfo {author} {\bibfnamefont {G.}~\bibnamefont {Vampa}},\ and\ \bibinfo
  {author} {\bibfnamefont {T.}~\bibnamefont {Brabec}},\ }\bibfield  {title}
  {\bibinfo {title} {Tunneling time, what is its meaning?},\ }
  \href@noop {}
  {\emph {\bibinfo {booktitle} {J. Phys.: Conf. Ser.}}}\
  \bibinfo {volume}\textbf{ {594}},\ 
   \bibinfo {pages} {012019}
   \bibinfo {year} ({2015})\
   \BibitemShut {NoStop}%
\bibitem [{\citenamefont {Pablico}\ and\ \citenamefont
  {Galapon}(2020)}]{20PablicoGalapon}%
  \BibitemOpen
  \bibfield  {author} {\bibinfo {author} {\bibfnamefont {D.~A.~L.}\
  \bibnamefont {Pablico}}\ and\ \bibinfo {author} {\bibfnamefont {E.~A.}\
  \bibnamefont {Galapon}},\ }\bibfield  {title} {\bibinfo {title} {Quantum
  traversal time across a potential well},\ }\href@noop {} {\bibfield
  {journal} {\bibinfo  {journal} {Phys. Rev. A.}\ }\textbf {\bibinfo {volume}
  {101}},\ \bibinfo {pages} {022103} (\bibinfo {year} {2020})}\BibitemShut
  {NoStop}%
\bibitem [{\citenamefont {Dumont}\ \emph {et~al.}(2020)\citenamefont {Dumont},
  \citenamefont {Rivlin},\ and\ \citenamefont {Pollak}}]{20DumontRivlinPollak}%
  \BibitemOpen
  \bibfield  {author} {\bibinfo {author} {\bibfnamefont {R.~S.}\ \bibnamefont
  {Dumont}}, \bibinfo {author} {\bibfnamefont {T.}~\bibnamefont {Rivlin}},\
  and\ \bibinfo {author} {\bibfnamefont {E.}~\bibnamefont {Pollak}},\
  }\bibfield  {title} {\bibinfo {title} {The relativistic tunneling flight time
  may be superluminal, but it does not imply superluminal signaling},\
  }\href@noop {} {\bibfield  {journal} {\bibinfo  {journal} {New J. Phys.}\
  }\textbf {\bibinfo {volume} {22}},\ \bibinfo {pages} {093060} (\bibinfo
  {year} {2020})}\BibitemShut {NoStop}%
\bibitem [{\citenamefont {Rivlin}\ \emph
  {et~al.}(2021{\natexlab{a}})\citenamefont {Rivlin}, \citenamefont {Pollak},\
  and\ \citenamefont {Dumont}}]{21RivlinPollakDumont}%
  \BibitemOpen
  \bibfield  {author} {\bibinfo {author} {\bibfnamefont {T.}~\bibnamefont
  {Rivlin}}, \bibinfo {author} {\bibfnamefont {E.}~\bibnamefont {Pollak}},\
  and\ \bibinfo {author} {\bibfnamefont {R.~S.}\ \bibnamefont {Dumont}},\
  }\bibfield  {title} {\bibinfo {title} {Determination of the tunneling flight
  time as the reflected phase time},\ }\href
  {https://doi.org/10.1103/PhysRevA.103.012225} {\bibfield  {journal} {\bibinfo
   {journal} {Phys. Rev. A}\ }\textbf {\bibinfo {volume} {103}},\ \bibinfo
  {pages} {012225} (\bibinfo {year} {2021}{\natexlab{a}})}\BibitemShut
  {NoStop}%
\bibitem [{\citenamefont {Rivlin}\ \emph
  {et~al.}(2021{\natexlab{b}})\citenamefont {Rivlin}, \citenamefont {Pollak},\
  and\ \citenamefont {Dumont}}]{21bRivlinPollakDumont}%
  \BibitemOpen
  \bibfield  {author} {\bibinfo {author} {\bibfnamefont {T.}~\bibnamefont
  {Rivlin}}, \bibinfo {author} {\bibfnamefont {E.}~\bibnamefont {Pollak}},\
  and\ \bibinfo {author} {\bibfnamefont {R.~S.}\ \bibnamefont {Dumont}},\
  }\bibfield  {title} {\bibinfo {title} {Comparison of a direct measure of
  barrier crossing times with indirect measures such as the larmor time},\
  }\href@noop {} {\bibfield  {journal} {\bibinfo  {journal} {New J. Phys.}\ }\textbf {\bibinfo {volume} {23}},\ \bibinfo
  {pages} {063044} (\bibinfo {year} {2021}{\natexlab{b}})}\BibitemShut {NoStop}%
\bibitem [{\citenamefont {Ianconescu}\ and\ \citenamefont
  {Pollak}(2021)}]{21IanconescuPollak}%
  \BibitemOpen
  \bibfield  {author} {\bibinfo {author} {\bibfnamefont {R.}~\bibnamefont
  {Ianconescu}}\ and\ \bibinfo {author} {\bibfnamefont {E.}~\bibnamefont
  {Pollak}},\ }\bibfield  {title} {\bibinfo {title} {Determination of the mean
  tunneling flight time in the b{\"u}ttiker-landauer oscillating-barrier model
  as the reflected phase time},\ }\href@noop {} {\bibfield  {journal} {\bibinfo
   {journal} {Phys. Rev. A}\ }\textbf {\bibinfo {volume} {103}},\ \bibinfo
  {pages} {042215} (\bibinfo {year} {2021})}\BibitemShut {NoStop}%
\bibitem [{\citenamefont {Sokolovski}\ and\ \citenamefont
  {Akhmatskaya}(2021)}]{21Sokolovski}%
  \BibitemOpen
  \bibfield  {author} {\bibinfo {author} {\bibfnamefont {D.}~\bibnamefont
  {Sokolovski}}\ and\ \bibinfo {author} {\bibfnamefont {E.}~\bibnamefont
  {Akhmatskaya}},\ }\bibfield  {title} {\bibinfo {title} {Tunnelling times, Larmor clock, and the elephant in the room},\ }\href@noop {} {\bibfield  {journal} {\bibinfo
   {journal} {Sci. Rep.}\ }\textbf {\bibinfo {volume} {11}},\ \bibinfo
  {pages} {10040} (\bibinfo {year} {2021})}\BibitemShut {NoStop}%
\bibitem [{\citenamefont {MacColl}(1932)}]{32Maccoll}%
  \BibitemOpen
  \bibfield  {author} {\bibinfo {author} {\bibfnamefont {L.}~\bibnamefont
  {MacColl}},\ }\bibfield  {title} {\bibinfo {title} {Note on the transmission
  and reflection of wave packets by potential barriers},\ }\href@noop {}
  {\bibfield  {journal} {\bibinfo  {journal} {Phys. Rev.}\ }\textbf {\bibinfo
  {volume} {40}},\ \bibinfo {pages} {621} (\bibinfo {year} {1932})}\BibitemShut
  {NoStop}%
\bibitem [{\citenamefont {Hartman}(1962)}]{62Hartman}%
  \BibitemOpen
  \bibfield  {author} {\bibinfo {author} {\bibfnamefont {T.~E.}\ \bibnamefont
  {Hartman}},\ }\bibfield  {title} {\bibinfo {title} {Tunneling of a wave
  packet},\ }\href@noop {} {\bibfield  {journal} {\bibinfo  {journal} {J. Appl.
  Phys.}\ }\textbf {\bibinfo {volume} {33}},\ \bibinfo {pages} {3427} (\bibinfo
  {year} {1962})}\BibitemShut {NoStop}%
\bibitem [{\citenamefont {Sainadh}\ \emph {et~al.}(2019)\citenamefont
  {Sainadh}, \citenamefont {Xu}, \citenamefont {Wang}, \citenamefont
  {Atia-Tul-Noor}, \citenamefont {Wallace}, \citenamefont {Douguet},
  \citenamefont {Bray}, \citenamefont {Ivanov}, \citenamefont {Bartschat},
  \citenamefont {Kheifets} \emph {et~al.}}]{19SainadhXu}%
  \BibitemOpen
  \bibfield  {author} {\bibinfo {author} {\bibfnamefont {U.~S.}\ \bibnamefont
  {Sainadh}}, \bibinfo {author} {\bibfnamefont {H.}~\bibnamefont {Xu}},
  \bibinfo {author} {\bibfnamefont {X.}~\bibnamefont {Wang}}, \bibinfo {author}
  {\bibfnamefont {A.}~\bibnamefont {Atia-Tul-Noor}}, \bibinfo {author}
  {\bibfnamefont {W.~C.}\ \bibnamefont {Wallace}}, \bibinfo {author}
  {\bibfnamefont {N.}~\bibnamefont {Douguet}}, \bibinfo {author} {\bibfnamefont
  {A.}~\bibnamefont {Bray}}, \bibinfo {author} {\bibfnamefont {I.}~\bibnamefont
  {Ivanov}}, \bibinfo {author} {\bibfnamefont {K.}~\bibnamefont {Bartschat}},
  \bibinfo {author} {\bibfnamefont {A.}~\bibnamefont {Kheifets}}, \emph
  {et~al.},\ }\bibfield  {title} {\bibinfo {title} {Attosecond angular
  streaking and tunnelling time in atomic hydrogen},\ }\href@noop {} {\bibfield
   {journal} {\bibinfo  {journal} {Nature}\ }\textbf {\bibinfo {volume}
  {568}},\ \bibinfo {pages} {75} (\bibinfo {year} {2019})}\BibitemShut
  {NoStop}%
\bibitem [{\citenamefont {Spierings}\ and\ \citenamefont
  {Steinberg}(2020)}]{20SpieringsSteinberg}%
  \BibitemOpen
  \bibfield  {author} {\bibinfo {author} {\bibfnamefont {D.~C.}\ \bibnamefont
  {Spierings}}\ and\ \bibinfo {author} {\bibfnamefont {A.~M.}\ \bibnamefont
  {Steinberg}},\ }\bibfield  {title} {\bibinfo {title} {Measuring the time
  tunneling particles spend in the barrier},\ }in\ \href@noop {} {\emph
  {\bibinfo {booktitle} {Optical, Opto-Atomic, and Entanglement-Enhanced
  Precision Metrology II}}},\ Vol.\ \bibinfo {volume} {11296}\ (\bibinfo
  {organization} {International Society for Optics and Photonics},\ \bibinfo
  {year} {2020})\ p.\ \bibinfo {pages} {112960F}\BibitemShut {NoStop}%
\bibitem [{\citenamefont {Spielmann}\ \emph {et~al.}(1994)\citenamefont
  {Spielmann}, \citenamefont {Szip{\"o}cs}, \citenamefont {Stingl},\ and\
  \citenamefont {Krausz}}]{94SpielmannSzipocs}%
  \BibitemOpen
  \bibfield  {author} {\bibinfo {author} {\bibfnamefont {C.}~\bibnamefont
  {Spielmann}}, \bibinfo {author} {\bibfnamefont {R.}~\bibnamefont
  {Szip{\"o}cs}}, \bibinfo {author} {\bibfnamefont {A.}~\bibnamefont
  {Stingl}},\ and\ \bibinfo {author} {\bibfnamefont {F.}~\bibnamefont
  {Krausz}},\ }\bibfield  {title} {\bibinfo {title} {Tunneling of Optical
  Pulses through Photonic Band Gaps},\ }\href@noop {} {\bibfield  {journal}
  {\bibinfo  {journal} {Phys. Rev. Lett.}\ }\textbf {\bibinfo {volume} {73}},\
  \bibinfo {pages} {2308} (\bibinfo {year} {1994})}\BibitemShut {NoStop}%
\bibitem [{\citenamefont {Diener}(1996)}]{96Diener}%
  \BibitemOpen
  \bibfield  {author} {\bibinfo {author} {\bibfnamefont {G.}~\bibnamefont
  {Diener}},\ }\bibfield  {title} {\bibinfo {title} {Superluminal group
  velocities and information transfer},\ }\href@noop {} {\bibfield  {journal}
  {\bibinfo  {journal} {Phys. Lett. A}\ }\textbf {\bibinfo {volume}
  {223}},\ \bibinfo {pages} {327} (\bibinfo {year} {1996})}\BibitemShut
  {NoStop}%
\bibitem [{\citenamefont {Chiao}\ and\ \citenamefont
  {Steinberg}(1997)}]{97ChiaoSteinberg}%
  \BibitemOpen
  \bibfield  {author} {\bibinfo {author} {\bibfnamefont {R.~Y.}\ \bibnamefont
  {Chiao}}\ and\ \bibinfo {author} {\bibfnamefont {A.~M.}\ \bibnamefont
  {Steinberg}},\ }\bibfield  {title} {\bibinfo {title} {Vi: tunneling times and
  superluminality},\ }in\ \href@noop {} {\emph {\bibinfo {booktitle} {Progress
  in Optics}}},\ Vol.~\bibinfo {volume} {37}\ (\bibinfo  {publisher}
  {Elsevier},\ \bibinfo {year} {1997})\ pp.\ \bibinfo {pages}
  {345--405}\BibitemShut {NoStop}%
\bibitem [{\citenamefont {Nimtz}(2003)}]{03Nimtz}%
  \BibitemOpen
  \bibfield  {author} {\bibinfo {author} {\bibfnamefont {G.}~\bibnamefont
  {Nimtz}},\ }\bibfield  {title} {\bibinfo {title} {On superluminal
  tunneling},\ }\href@noop {} {\bibfield  {journal} {\bibinfo  {journal}
  {Prog. Quantum Electron.}\ }\textbf {\bibinfo {volume} {27}},\
  \bibinfo {pages} {417} (\bibinfo {year} {2003})}\BibitemShut {NoStop}%
\bibitem [{\citenamefont {Winful}(2006{\natexlab{a}})}]{06Winful}%
  \BibitemOpen
  \bibfield  {author} {\bibinfo {author} {\bibfnamefont {H.~G.}\ \bibnamefont
  {Winful}},\ }\bibfield  {title} {\bibinfo {title} {Tunneling time, the
  hartman effect, and superluminality: A proposed resolution of an old
  paradox},\ }\href@noop {} {\bibfield  {journal} {\bibinfo  {journal} {Phys.
  Rep.}\ }\textbf {\bibinfo {volume} {436}},\ \bibinfo {pages} {1} (\bibinfo
  {year} {2006}{\natexlab{a}})}\BibitemShut {NoStop}%
\bibitem [{\citenamefont {Winful}(2006{\natexlab{b}})}]{06Winfulb}%
  \BibitemOpen
  \bibfield  {author} {\bibinfo {author} {\bibfnamefont {H.~G.}\ \bibnamefont
  {Winful}},\ }\bibfield  {title} {\bibinfo {title} {The meaning of group delay
  in barrier tunnelling: a re-examination of superluminal group velocities},\
  }\href@noop {} {\bibfield  {journal} {\bibinfo  {journal} {New J. Phys.}\
  }\textbf {\bibinfo {volume} {8}},\ \bibinfo {pages} {101} (\bibinfo {year}
  {2006}{\natexlab{b}})}\BibitemShut {NoStop}%
\bibitem [{\citenamefont {Nimtz}\ and\ \citenamefont
  {Aichmann}(2017)}]{17NimtzAichmann}%
  \BibitemOpen
  \bibfield  {author} {\bibinfo {author} {\bibfnamefont {G.}~\bibnamefont
  {Nimtz}}\ and\ \bibinfo {author} {\bibfnamefont {H.}~\bibnamefont
  {Aichmann}},\ }\bibfield  {title} {\bibinfo {title} {Zero-time
  tunneling--revisited},\ }\href@noop {} {\bibfield  {journal} {\bibinfo
  {journal} {Z. Naturforsch. A}\ }\textbf {\bibinfo {volume}
  {72}},\ \bibinfo {pages} {881} (\bibinfo {year} {2017})}\BibitemShut
  {NoStop}%
\bibitem [{\citenamefont {Dumont}\ and\ \citenamefont
  {Marchioro~II}(1993)}]{93DumontMarchioro}%
  \BibitemOpen
  \bibfield  {author} {\bibinfo {author} {\bibfnamefont {R.~S.}\ \bibnamefont
  {Dumont}}\ and\ \bibinfo {author} {\bibfnamefont {T.~L.}~\bibnamefont
  {Marchioro~II}},\ }\bibfield  {title} {\bibinfo {title} {Tunneling-time
  probability distribution},\ }\href@noop {} {\bibfield  {journal} {\bibinfo
  {journal} {Phys. Rev. A.}\ }\textbf {\bibinfo {volume} {47}},\ \bibinfo
  {pages} {85} (\bibinfo {year} {1993})}\BibitemShut {NoStop}%
\bibitem [{\citenamefont {De~Leo}\ and\ \citenamefont
  {Rotelli}(2007)}]{07DeLeoRotelli}%
  \BibitemOpen
  \bibfield  {author} {\bibinfo {author} {\bibfnamefont {S.}~\bibnamefont
  {De~Leo}}\ and\ \bibinfo {author} {\bibfnamefont {P.~P.}\ \bibnamefont
  {Rotelli}},\ }\bibfield  {title} {\bibinfo {title} {Dirac equation studies in
  the tunneling energy zone},\ }\href@noop {} {\bibfield  {journal} {\bibinfo
  {journal} {Eur. Phys. J. C}\ }\textbf {\bibinfo {volume}
  {51}},\ \bibinfo {pages} {241} (\bibinfo {year} {2007})}\BibitemShut
  {NoStop}%
\bibitem [{\citenamefont {Fuhrmanek}\ \emph {et~al.}(2010)\citenamefont
  {Fuhrmanek}, \citenamefont {Lance}, \citenamefont {Tuchendler}, \citenamefont
  {Grangier}, \citenamefont {Sortais},\ and\ \citenamefont
  {Browaeys}}]{10FuhrmanekLanceTuchendler}%
  \BibitemOpen
  \bibfield  {author} {\bibinfo {author} {\bibfnamefont {A.}~\bibnamefont
  {Fuhrmanek}}, \bibinfo {author} {\bibfnamefont {A.~M.}\ \bibnamefont
  {Lance}}, \bibinfo {author} {\bibfnamefont {C.}~\bibnamefont {Tuchendler}},
  \bibinfo {author} {\bibfnamefont {P.}~\bibnamefont {Grangier}}, \bibinfo
  {author} {\bibfnamefont {Y.~R.}\ \bibnamefont {Sortais}},\ and\ \bibinfo
  {author} {\bibfnamefont {A.}~\bibnamefont {Browaeys}},\ }\bibfield  {title}
  {\bibinfo {title} {Imaging a single atom in a time-of-flight experiment},\
  }\href@noop {} {\bibfield  {journal} {\bibinfo  {journal} {New J. Phys.}\
  }\textbf {\bibinfo {volume} {12}},\ \bibinfo {pages} {053028} (\bibinfo
  {year} {2010})}\BibitemShut {NoStop}%
\bibitem [{\citenamefont {Du}\ \emph {et~al.}(2015)\citenamefont {Du},
  \citenamefont {Li}, \citenamefont {Wen}, \citenamefont {Li},\ and\
  \citenamefont {Zhang}}]{15DuLiWen}%
  \BibitemOpen
  \bibfield  {author} {\bibinfo {author} {\bibfnamefont {J.-J.}\ \bibnamefont
  {Du}}, \bibinfo {author} {\bibfnamefont {W.-F.}\ \bibnamefont {Li}}, \bibinfo
  {author} {\bibfnamefont {R.-J.}\ \bibnamefont {Wen}}, \bibinfo {author}
  {\bibfnamefont {G.}~\bibnamefont {Li}},\ and\ \bibinfo {author}
  {\bibfnamefont {T.-C.}\ \bibnamefont {Zhang}},\ }\bibfield  {title} {\bibinfo
  {title} {Experimental investigation of the statistical distribution of single
  atoms in cavity quantum electrodynamics},\ }\href@noop {} {\bibfield
  {journal} {\bibinfo  {journal} {Laser Phys. Lett.}\ }\textbf {\bibinfo
  {volume} {12}},\ \bibinfo {pages} {065501} (\bibinfo {year}
  {2015})}\BibitemShut {NoStop}%
\bibitem [{\citenamefont {Ramos}\ \emph {et~al.}(2020)\citenamefont {Ramos},
  \citenamefont {Spierings}, \citenamefont {Racicot},\ and\ \citenamefont
  {Steinberg}}]{20RamosSpierings}%
  \BibitemOpen
  \bibfield  {author} {\bibinfo {author} {\bibfnamefont {R.}~\bibnamefont
  {Ramos}}, \bibinfo {author} {\bibfnamefont {D.}~\bibnamefont {Spierings}},
  \bibinfo {author} {\bibfnamefont {I.}~\bibnamefont {Racicot}},\ and\ \bibinfo
  {author} {\bibfnamefont {A.}~\bibnamefont {Steinberg}},\ }\bibfield  {title}
  {\bibinfo {title} {Measurement of the time spent by a tunnelling atom within
  the barrier region},\ }\href@noop {} {\bibfield  {journal} {\bibinfo
  {journal} {Nature}\ }\textbf {\bibinfo {volume} {583}},\ \bibinfo {pages}
  {529} (\bibinfo {year} {2020})}\BibitemShut {NoStop}%
\bibitem [{\citenamefont {Spierings}\ and\ \citenamefont
  {Steinberg}(2021)}]{21SpieringsSteinberg}%
  \BibitemOpen
  \bibfield  {author} {\bibinfo {author} {\bibfnamefont {D.~C.}\ \bibnamefont
  {Spierings}}\ and\ \bibinfo {author} {\bibfnamefont {A.~M.}\ \bibnamefont
  {Steinberg}},\ }\bibfield  {title} {\bibinfo {title} {Observation of the
  Decrease of Larmor Tunneling Times with Lower Incident Energy},\ }\href@noop
  {} {\bibfield  {journal} {\bibinfo  {journal} {Phys. Rev. Lett.}\
  }\textbf {\bibinfo {volume} {127}},\ \bibinfo {pages} {133001} (\bibinfo
  {year} {2021})}\BibitemShut {NoStop}%
\bibitem [{\citenamefont {Steinberg}\ \emph {et~al.}(1993)\citenamefont
  {Steinberg}, \citenamefont {Kwiat},\ and\ \citenamefont
  {Chiao}}]{93SteinbergKwiatChiao}%
  \BibitemOpen
  \bibfield  {author} {\bibinfo {author} {\bibfnamefont {A.~M.}\ \bibnamefont
  {Steinberg}}, \bibinfo {author} {\bibfnamefont {P.~G.}\ \bibnamefont
  {Kwiat}},\ and\ \bibinfo {author} {\bibfnamefont {R.~Y.}\ \bibnamefont
  {Chiao}},\ }\bibfield  {title} {\bibinfo {title} {Measurement of the
  single-photon tunneling time},\ }\href@noop {} {\bibfield  {journal}
  {\bibinfo  {journal} {Phys. Rev. Lett.}\ }\textbf {\bibinfo {volume} {71}},\
  \bibinfo {pages} {708} (\bibinfo {year} {1993})}\BibitemShut {NoStop}%
\bibitem [{\citenamefont {Longhi}\ \emph {et~al.}(2001)\citenamefont {Longhi},
  \citenamefont {Marano}, \citenamefont {Laporta},\ and\ \citenamefont
  {Belmonte}}]{01LonghiMarano}%
  \BibitemOpen
  \bibfield  {author} {\bibinfo {author} {\bibfnamefont {S.}~\bibnamefont
  {Longhi}}, \bibinfo {author} {\bibfnamefont {M.}~\bibnamefont {Marano}},
  \bibinfo {author} {\bibfnamefont {P.}~\bibnamefont {Laporta}},\ and\ \bibinfo
  {author} {\bibfnamefont {M.}~\bibnamefont {Belmonte}},\ }\bibfield  {title}
  {\bibinfo {title} {Superluminal optical pulse propagation at 1.5 $\mu$ m in
  periodic fiber bragg gratings},\ }\href@noop {} {\bibfield  {journal}
  {\bibinfo  {journal} {Phys. Rev. E}\ }\textbf {\bibinfo {volume} {64}},\
  \bibinfo {pages} {055602(R)} (\bibinfo {year} {2001})}\BibitemShut {NoStop}%
\bibitem [{\citenamefont {Shafir}\ \emph {et~al.}(2012)\citenamefont {Shafir},
  \citenamefont {Soifer}, \citenamefont {Bruner}, \citenamefont {Dagan},
  \citenamefont {Mairesse}, \citenamefont {Patchkovskii}, \citenamefont
  {Ivanov}, \citenamefont {Smirnova},\ and\ \citenamefont
  {Dudovich}}]{12ShafirSoiferBruner}%
  \BibitemOpen
  \bibfield  {author} {\bibinfo {author} {\bibfnamefont {D.}~\bibnamefont
  {Shafir}}, \bibinfo {author} {\bibfnamefont {H.}~\bibnamefont {Soifer}},
  \bibinfo {author} {\bibfnamefont {B.~D.}\ \bibnamefont {Bruner}}, \bibinfo
  {author} {\bibfnamefont {M.}~\bibnamefont {Dagan}}, \bibinfo {author}
  {\bibfnamefont {Y.}~\bibnamefont {Mairesse}}, \bibinfo {author}
  {\bibfnamefont {S.}~\bibnamefont {Patchkovskii}}, \bibinfo {author}
  {\bibfnamefont {M.~Y.}\ \bibnamefont {Ivanov}}, \bibinfo {author}
  {\bibfnamefont {O.}~\bibnamefont {Smirnova}},\ and\ \bibinfo {author}
  {\bibfnamefont {N.}~\bibnamefont {Dudovich}},\ }\bibfield  {title} {\bibinfo
  {title} {Resolving the time when an electron exits a tunnelling barrier},\
  }\href@noop {} {\bibfield  {journal} {\bibinfo  {journal} {Nature}\ }\textbf
  {\bibinfo {volume} {485}},\ \bibinfo {pages} {343} (\bibinfo {year}
  {2012})}\BibitemShut {NoStop}%
\bibitem [{\citenamefont {Landsman}\ \emph {et~al.}(2014)\citenamefont
  {Landsman}, \citenamefont {Weger}, \citenamefont {Maurer}, \citenamefont
  {Boge}, \citenamefont {Ludwig}, \citenamefont {Heuser}, \citenamefont
  {Cirelli}, \citenamefont {Gallmann},\ and\ \citenamefont
  {Keller}}]{14LandsmanWeger}%
  \BibitemOpen
  \bibfield  {author} {\bibinfo {author} {\bibfnamefont {A.~S.}\ \bibnamefont
  {Landsman}}, \bibinfo {author} {\bibfnamefont {M.}~\bibnamefont {Weger}},
  \bibinfo {author} {\bibfnamefont {J.}~\bibnamefont {Maurer}}, \bibinfo
  {author} {\bibfnamefont {R.}~\bibnamefont {Boge}}, \bibinfo {author}
  {\bibfnamefont {A.}~\bibnamefont {Ludwig}}, \bibinfo {author} {\bibfnamefont
  {S.}~\bibnamefont {Heuser}}, \bibinfo {author} {\bibfnamefont
  {C.}~\bibnamefont {Cirelli}}, \bibinfo {author} {\bibfnamefont
  {L.}~\bibnamefont {Gallmann}},\ and\ \bibinfo {author} {\bibfnamefont
  {U.}~\bibnamefont {Keller}},\ }\bibfield  {title} {\bibinfo {title}
  {Ultrafast resolution of tunneling delay time},\ }\href@noop {} {\bibfield
  {journal} {\bibinfo  {journal} {Optica}\ }\textbf {\bibinfo {volume} {1}},\
  \bibinfo {pages} {343} (\bibinfo {year} {2014})}\BibitemShut {NoStop}%
\bibitem [{\citenamefont {Landsman}\ and\ \citenamefont
  {Keller}(2015)}]{15LandsmanKeller}%
  \BibitemOpen
  \bibfield  {author} {\bibinfo {author} {\bibfnamefont {A.~S.}\ \bibnamefont
  {Landsman}}\ and\ \bibinfo {author} {\bibfnamefont {U.}~\bibnamefont
  {Keller}},\ }\bibfield  {title} {\bibinfo {title} {Attosecond science and the
  tunnelling time problem},\ }\href@noop {} {\bibfield  {journal} {\bibinfo
  {journal} {Phys. Rep.}\ }\textbf {\bibinfo {volume} {547}},\ \bibinfo {pages}
  {1} (\bibinfo {year} {2015})}\BibitemShut {NoStop}%
\bibitem [{\citenamefont {Torlina}\ \emph {et~al.}(2015)\citenamefont
  {Torlina}, \citenamefont {Morales}, \citenamefont {Kaushal}, \citenamefont
  {Ivanov}, \citenamefont {Kheifets}, \citenamefont {Zielinski}, \citenamefont
  {Scrinzi}, \citenamefont {Muller}, \citenamefont {Sukiasyan}, \citenamefont
  {Ivanov} \emph {et~al.}}]{15TorlinaMorales}%
  \BibitemOpen
  \bibfield  {author} {\bibinfo {author} {\bibfnamefont {L.}~\bibnamefont
  {Torlina}}, \bibinfo {author} {\bibfnamefont {F.}~\bibnamefont {Morales}},
  \bibinfo {author} {\bibfnamefont {J.}~\bibnamefont {Kaushal}}, \bibinfo
  {author} {\bibfnamefont {I.}~\bibnamefont {Ivanov}}, \bibinfo {author}
  {\bibfnamefont {A.}~\bibnamefont {Kheifets}}, \bibinfo {author}
  {\bibfnamefont {A.}~\bibnamefont {Zielinski}}, \bibinfo {author}
  {\bibfnamefont {A.}~\bibnamefont {Scrinzi}}, \bibinfo {author} {\bibfnamefont
  {H.~G.}\ \bibnamefont {Muller}}, \bibinfo {author} {\bibfnamefont
  {S.}~\bibnamefont {Sukiasyan}}, \bibinfo {author} {\bibfnamefont
  {M.}~\bibnamefont {Ivanov}}, \emph {et~al.},\ }\bibfield  {title} {\bibinfo
  {title} {Interpreting attoclock measurements of tunnelling times},\
  }\href@noop {} {\bibfield  {journal} {\bibinfo  {journal} {Nat. Phys.}\
  }\textbf {\bibinfo {volume} {11}},\ \bibinfo {pages} {503} (\bibinfo {year}
  {2015})}\BibitemShut {NoStop}%
\bibitem [{\citenamefont {Pedatzur}\ \emph {et~al.}(2015)\citenamefont
  {Pedatzur}, \citenamefont {Orenstein}, \citenamefont {Serbinenko},
  \citenamefont {Soifer}, \citenamefont {Bruner}, \citenamefont {Uzan},
  \citenamefont {Brambila}, \citenamefont {Harvey}, \citenamefont {Torlina},
  \citenamefont {Morales} \emph {et~al.}}]{15PedatzurOrensteinSerbinenko}%
  \BibitemOpen
  \bibfield  {author} {\bibinfo {author} {\bibfnamefont {O.}~\bibnamefont
  {Pedatzur}}, \bibinfo {author} {\bibfnamefont {G.}~\bibnamefont {Orenstein}},
  \bibinfo {author} {\bibfnamefont {V.}~\bibnamefont {Serbinenko}}, \bibinfo
  {author} {\bibfnamefont {H.}~\bibnamefont {Soifer}}, \bibinfo {author}
  {\bibfnamefont {B.}~\bibnamefont {Bruner}}, \bibinfo {author} {\bibfnamefont
  {A.}~\bibnamefont {Uzan}}, \bibinfo {author} {\bibfnamefont {D.}~\bibnamefont
  {Brambila}}, \bibinfo {author} {\bibfnamefont {A.}~\bibnamefont {Harvey}},
  \bibinfo {author} {\bibfnamefont {L.}~\bibnamefont {Torlina}}, \bibinfo
  {author} {\bibfnamefont {F.}~\bibnamefont {Morales}}, \emph {et~al.},\
  }\bibfield  {title} {\bibinfo {title} {Attosecond tunnelling
  interferometry},\ }\href@noop {} {\bibfield  {journal} {\bibinfo  {journal}
  {Nat. Phys.}\ }\textbf {\bibinfo {volume} {11}},\ \bibinfo {pages} {815}
  (\bibinfo {year} {2015})}\BibitemShut {NoStop}%
\bibitem [{\citenamefont {Texier}(2016)}]{16Texier}%
  \BibitemOpen
  \bibfield  {author} {\bibinfo {author} {\bibfnamefont {C.}~\bibnamefont
  {Texier}},\ }\bibfield  {title} {\bibinfo {title} {Wigner time delay and
  related concepts: Application to transport in coherent conductors},\
  }\href@noop {} {\bibfield  {journal} {\bibinfo  {journal} {Phys. E}\
  }\textbf {\bibinfo {volume} {82}},\ \bibinfo {pages} {16} (\bibinfo {year}
  {2016})}\BibitemShut {NoStop}%
\bibitem [{\citenamefont {Sainadh}\ \emph {et~al.}(2020)\citenamefont
  {Sainadh}, \citenamefont {Sang},\ and\ \citenamefont
  {Litvinyuk}}]{20SainadhSangLitvinyuk}%
  \BibitemOpen
  \bibfield  {author} {\bibinfo {author} {\bibfnamefont {S.~U.}\ \bibnamefont
  {Sainadh}}, \bibinfo {author} {\bibfnamefont {R.~T.}\ \bibnamefont {Sang}},\
  and\ \bibinfo {author} {\bibfnamefont {I.~V.}\ \bibnamefont {Litvinyuk}},\
  }\bibfield  {title} {\bibinfo {title} {Attoclock and the quest for tunnelling
  time in strong-field physics},\ }\href@noop {} {\bibfield  {journal}
  {\bibinfo  {journal} {J. Phys. Photon.}\ }\textbf {\bibinfo {volume} {2}},\
  \bibinfo {pages} {042002} (\bibinfo {year} {2020})}\BibitemShut {NoStop}%
\bibitem [{\citenamefont {Kheifets}(2020)}]{20Kheifets}%
  \BibitemOpen
  \bibfield  {author} {\bibinfo {author} {\bibfnamefont {A.~S.}\ \bibnamefont
  {Kheifets}},\ }\bibfield  {title} {\bibinfo {title} {The attoclock and the
  tunneling time debate},\ }\href@noop {} {\bibfield  {journal} {\bibinfo
  {journal} {J. Phys. B-At. Mol. Opt.}\ }\textbf {\bibinfo {volume} {53}},\
  \bibinfo {pages} {072001} (\bibinfo {year} {2020})}\BibitemShut {NoStop}%
\bibitem [{\citenamefont {Petersen}\ and\ \citenamefont
  {Pollak}(2017)}]{17PetersenPollak}%
  \BibitemOpen
  \bibfield  {author} {\bibinfo {author} {\bibfnamefont {J.}~\bibnamefont
  {Petersen}}\ and\ \bibinfo {author} {\bibfnamefont {E.}~\bibnamefont
  {Pollak}},\ }\bibfield  {title} {\bibinfo {title} {Tunneling flight time,
  chemistry, and special relativity},\ }\href@noop {} {\bibfield  {journal}
  {\bibinfo  {journal} {J. Phys. Chem. Lett.}\ }\textbf {\bibinfo {volume}
  {8}},\ \bibinfo {pages} {4017} (\bibinfo {year} {2017})}\BibitemShut
  {NoStop}%
\bibitem [{\citenamefont {Pollak}(2017)}]{17Pollak}%
  \BibitemOpen
  \bibfield  {author} {\bibinfo {author} {\bibfnamefont {E.}~\bibnamefont
  {Pollak}},\ }\bibfield  {title} {\bibinfo {title} {Transition Path Time
  Distribution, Tunneling Times, Friction, and Uncertainty},\ }\href@noop {}
  {\bibfield  {journal} {\bibinfo  {journal} {Phys. Rev. Lett.}\ }\textbf
  {\bibinfo {volume} {118}},\ \bibinfo {pages} {070401} (\bibinfo {year}
  {2017})}\BibitemShut {NoStop}%
\bibitem [{\citenamefont {Pollak}\ and\ \citenamefont
  {Miret-Art{\'e}s}(2018)}]{18PollakMiret}%
  \BibitemOpen
  \bibfield  {author} {\bibinfo {author} {\bibfnamefont {E.}~\bibnamefont
  {Pollak}}\ and\ \bibinfo {author} {\bibfnamefont {S.}~\bibnamefont
  {Miret-Art{\'e}s}},\ }\bibfield  {title} {\bibinfo {title} {Time averaging of
  weak values—consequences for time-energy and coordinate-momentum
  uncertainty},\ }\href@noop {} {\bibfield  {journal} {\bibinfo  {journal} {New
  J. Phys.}\ }\textbf {\bibinfo {volume} {20}},\ \bibinfo {pages} {073016}
  (\bibinfo {year} {2018})}\BibitemShut {NoStop}%
\bibitem [{\citenamefont {Li}\ and\ \citenamefont {Chen}(2002)}]{02LiChen}%
  \BibitemOpen
  \bibfield  {author} {\bibinfo {author} {\bibfnamefont {C.-F.}\ \bibnamefont
  {Li}}\ and\ \bibinfo {author} {\bibfnamefont {X.}~\bibnamefont {Chen}},\
  }\bibfield  {title} {\bibinfo {title} {Traversal time for dirac particles
  through a potential barrier},\ }\href@noop {} {\bibfield  {journal} {\bibinfo
   {journal} {Ann. Phys. (Leipzig)}\ }\textbf {\bibinfo {volume} {11}},\ \bibinfo
  {pages} {916} (\bibinfo {year} {2002})}\BibitemShut {NoStop}%
\bibitem [{\citenamefont {Lunardi}\ and\ \citenamefont
  {Manzoni}(2007)}]{07LunardiManzoni}%
  \BibitemOpen
  \bibfield  {author} {\bibinfo {author} {\bibfnamefont {J.~T.}\ \bibnamefont
  {Lunardi}}\ and\ \bibinfo {author} {\bibfnamefont {L.~A.}\ \bibnamefont
  {Manzoni}},\ }\bibfield  {title} {\bibinfo {title} {Relativistic tunneling
  through two successive barriers},\ }\href@noop {} {\bibfield  {journal}
  {\bibinfo  {journal} {Phys. Rev. A.}\ }\textbf {\bibinfo {volume} {76}},\
  \bibinfo {pages} {042111} (\bibinfo {year} {2007})}\BibitemShut {NoStop}%
\bibitem [{\citenamefont {De~Leo}(2013)}]{13DeLeo}%
  \BibitemOpen
  \bibfield  {author} {\bibinfo {author} {\bibfnamefont {S.}~\bibnamefont
  {De~Leo}},\ }\bibfield  {title} {\bibinfo {title} {A study of transit times
  in dirac tunneling},\ }\href@noop {} {\bibfield  {journal} {\bibinfo
  {journal} {J. of Phys. A: Math. Theor.}\ }\textbf
  {\bibinfo {volume} {46}},\ \bibinfo {pages} {155306} (\bibinfo {year}
  {2013})}\BibitemShut {NoStop}%
\bibitem [{\citenamefont {Ivanov}\ and\ \citenamefont
  {Kim}(2015)}]{15IvanovKim}%
  \BibitemOpen
  \bibfield  {author} {\bibinfo {author} {\bibfnamefont {I.~A.}~\bibnamefont
  {Ivanov}}\ and\ \bibinfo {author} {\bibfnamefont {K.~T.}\ \bibnamefont
  {Kim}},\ }\bibfield  {title} {\bibinfo {title} {Relativistic approach to the
  tunneling-time problem},\ }\href@noop {} {\bibfield  {journal} {\bibinfo
  {journal} {Phys. Rev. A.}\ }\textbf {\bibinfo {volume} {92}},\ \bibinfo
  {pages} {053418} (\bibinfo {year} {2015})}\BibitemShut {NoStop}%
\bibitem [{\citenamefont {Nitta}\ \emph {et~al.}(1999)\citenamefont {Nitta},
  \citenamefont {Kudo},\ and\ \citenamefont {Minowa}}]{99NittaKudoMinowa}%
  \BibitemOpen
  \bibfield  {author} {\bibinfo {author} {\bibfnamefont {H.}~\bibnamefont
  {Nitta}}, \bibinfo {author} {\bibfnamefont {T.}~\bibnamefont {Kudo}},\ and\
  \bibinfo {author} {\bibfnamefont {H.}~\bibnamefont {Minowa}},\ }\bibfield
  {title} {\bibinfo {title} {Motion of a wave packet in the klein paradox},\
  }\href@noop {} {\bibfield  {journal} {\bibinfo  {journal} {Am. J. Phys.}\
  }\textbf {\bibinfo {volume} {67}},\ \bibinfo {pages} {966} (\bibinfo {year}
  {1999})}\BibitemShut {NoStop}%
\bibitem [{\citenamefont {Asfaw}\ \emph {et~al.}(2013)\citenamefont {Asfaw},
  \citenamefont {Alvarez-Lacalle},\ and\ \citenamefont {Shiferaw}}]{13AsfawAlvarezLacalleShiferaw}%
  \BibitemOpen
  \bibfield  {author} {\bibinfo {author} {\bibfnamefont {M.}~\bibnamefont
  {Asfaw}}, \bibinfo {author} {\bibfnamefont {E.}~\bibnamefont {Alvarez-Lacalle}},\ and\
  \bibinfo {author} {\bibfnamefont {Y.}~\bibnamefont {Shiferaw}},\ }\bibfield
  {title} {\bibinfo {title} {The Timing Statistics of Spontaneous Calcium Release in Cardiac Myocytes},\
  }\href@noop {} {\bibfield  {journal} {\bibinfo  {journal} {PLoS One}\
  }\textbf {\bibinfo {volume} {8}},\ \bibinfo {pages} {e62967} (\bibinfo {year}
  {2013})}\BibitemShut {NoStop}%
\bibitem [{\citenamefont {B{\"u}ttiker}\ and\ \citenamefont
  {Washburn}(2003)}]{03ButtikerWashburn}%
  \BibitemOpen
  \bibfield  {author} {\bibinfo {author} {\bibfnamefont {M.}~\bibnamefont
  {B{\"u}ttiker}}\ and\ \bibinfo {author} {\bibfnamefont {S.}~\bibnamefont
  {Washburn}},\ }\bibfield  {title} {\bibinfo {title} {Ado about nothing
  much?},\ }\href@noop {} {\bibfield  {journal} {\bibinfo  {journal} {Nature}\
  }\textbf {\bibinfo {volume} {422}},\ \bibinfo {pages} {271} (\bibinfo {year}
  {2003})}\BibitemShut {NoStop}%
\bibitem [{\citenamefont {Bracher}\ \emph {et~al.}(1998)\citenamefont
  {Bracher}, \citenamefont {Becker}, \citenamefont {Gurvitz}, \citenamefont
  {Kleber},\ and\ \citenamefont {Marinov}}]{98BracherBeckerGurvitz}%
  \BibitemOpen
  \bibfield  {author} {\bibinfo {author} {\bibfnamefont {C.}~\bibnamefont
  {Bracher}}, \bibinfo {author} {\bibfnamefont {W.}~\bibnamefont {Becker}},
  \bibinfo {author} {\bibfnamefont {S.}~\bibnamefont {Gurvitz}}, \bibinfo
  {author} {\bibfnamefont {M.}~\bibnamefont {Kleber}},\ and\ \bibinfo {author}
  {\bibfnamefont {M.}~\bibnamefont {Marinov}},\ }\bibfield  {title} {\bibinfo
  {title} {Three-dimensional tunneling in quantum ballistic motion},\
  }\href@noop {} {\bibfield  {journal} {\bibinfo  {journal} {Am. J. Phys.}\
  }\textbf {\bibinfo {volume} {66}},\ \bibinfo {pages} {38} (\bibinfo {year}
  {1998})}\BibitemShut {NoStop}%
\bibitem [{\citenamefont {Nandi}\ \emph {et~al.}(2023)\citenamefont {Nandi},
  \citenamefont {Laude}, \citenamefont {Khire}, \citenamefont {Gurav},
  \citenamefont {Qu}, \citenamefont {Conte}, \citenamefont {Yu}, \citenamefont
  {Li}, \citenamefont {Houston}, \citenamefont {Gadre} \emph
  {et~al.}}]{23NandiLaudeKhire}%
  \BibitemOpen
  \bibfield  {author} {\bibinfo {author} {\bibfnamefont {A.}~\bibnamefont
  {Nandi}}, \bibinfo {author} {\bibfnamefont {G.}~\bibnamefont {Laude}},
  \bibinfo {author} {\bibfnamefont {S.~S.}\ \bibnamefont {Khire}}, \bibinfo
  {author} {\bibfnamefont {N.~D.}\ \bibnamefont {Gurav}}, \bibinfo {author}
  {\bibfnamefont {C.}~\bibnamefont {Qu}}, \bibinfo {author} {\bibfnamefont
  {R.}~\bibnamefont {Conte}}, \bibinfo {author} {\bibfnamefont
  {Q.}~\bibnamefont {Yu}}, \bibinfo {author} {\bibfnamefont {S.}~\bibnamefont
  {Li}}, \bibinfo {author} {\bibfnamefont {P.~L.}\ \bibnamefont {Houston}},
  \bibinfo {author} {\bibfnamefont {S.~R.}\ \bibnamefont {Gadre}}, \emph
  {et~al.},\ }\bibfield  {title} {\bibinfo {title} {Ring-polymer instanton
  tunneling splittings of tropolone and isotopomers using a $\delta$-machine
  learned ccsd (t) potential: Theory and experiment shake hands},\ }\href
  {https://doi.org/10.1021/jacs.3c00769} {\bibfield  {journal} {\bibinfo
  {journal} {J. Am. Chem. Soc.}\ \textbf {\bibinfo {volume} {145}}, \bibinfo {pages} {9655}} (\bibinfo
  {year} {2023})}\BibitemShut {NoStop}%
\end{thebibliography}

\providecommand{\noopsort}[1]{}\providecommand{\singleletter}[1]{#1}%

\end{document}